\documentclass[manuscript,screen,nonacm]{acmart}

\usepackage{longtable}
\usepackage{capt-of}
\usepackage{balance,bm}
\usepackage{color, colortbl, xcolor}
\usepackage{todonotes}
\definecolor{LightGray}{gray}{0.97}
\usepackage{pgfplots}
\usepackage{xcolor}
\usepackage{url}
\usepackage{tabularx}
\usepackage{float}

\usepackage{subcaption}
\usepackage{booktabs} 
\usepackage{graphicx}
\usepackage{textcomp}
\usepackage{color,soul}
\usepackage{bm}
\usepackage{multirow}
\usepackage{wrapfig}
\usepackage{balance}
\usepackage{graphicx}  
\usepackage{enumitem}

\usepackage{colortbl}
\usepackage{arydshln}
\usepackage [english]{babel}
\usepackage [autostyle, english = american]{csquotes}
\definecolor{linkColor}{RGB}{6,125,233}
\definecolor{green}{rgb}{0.0, 0.65, 0.31}
\definecolor{bleudefrance}{rgb}{0.19, 0.55, 0.91}
\definecolor{ceruleanblue}{rgb}{0.16, 0.32, 0.75}
\definecolor{grey}{HTML}{969696}
\definecolor{violet}{HTML}{756bb1}
\definecolor{dgrey}{HTML}{01665e}
\definecolor{lgrey}{HTML}{5ab4ac}
\definecolor{dgreen}{HTML}{005a32}
\definecolor{purple}{HTML}{ae017e}

\definecolor{editCol}{HTML}{000000}
\definecolor{maskCol}{HTML}{c51b7d}
\definecolor{lrColor}{HTML}{8856a7}
\definecolor{trColor}{HTML}{d01c8b}
\definecolor{ctColor}{HTML}{4dac26}
\definecolor{brickred}{HTML}{f03b20}
\definecolor{DarkBlue}{HTML}{00008B}
\definecolor{mscolor}{HTML}{01665e}
\definecolor{nmscolor}{HTML}{bf812d}
\definecolor{lgreen}{HTML}{ccece6}
\definecolor{dolive}{HTML}{308014}

\definecolor{maskCol}{HTML}{c51b7d}
\definecolor{lrColor}{HTML}{8856a7}
\definecolor{trColor}{HTML}{d01c8b}
\definecolor{ctColor}{HTML}{4dac26}
\definecolor{brickred}{HTML}{f03b20}
\definecolor{lgreen}{HTML}{e0f3db}
\definecolor{dpink}{HTML}{CD1076}
\definecolor{pink}{HTML}{FED2D2}
\definecolor{soothinggreen}{HTML}{4dac26}
\definecolor{darkred}{HTML}{8B0000}

\definecolor{dblue}{HTML}{215F9A}
\definecolor{violet}{HTML}{8A2BE2}
\definecolor{mscolor}{HTML}{01665e}
\definecolor{nmscolor}{HTML}{d8b365}
\definecolor{deepgrey}{HTML}{525252}
\definecolor{dslate}{HTML}{2F4F4F}
\definecolor{dolive}{HTML}{556B2F}
\definecolor{teal}{HTML}{388E8E}
\definecolor{mscolor}{HTML}{01665e}
\definecolor{nmscolor}{HTML}{d8b365}

\definecolor{aicolor}{HTML}{018571}
\definecolor{occolor}{HTML}{ff7799}

\definecolor{srcolor}{HTML}{e34a33}
\definecolor{smcolor}{HTML}{253494}
\definecolor{srsmcolor}{HTML}{7fcdbb}
\definecolor{bothcolor}{HTML}{fe9929}
\definecolor{onecolor}{HTML}{018571}
\definecolor{marroon}{HTML}{881c1c}

\usepackage{mathtools}

\usepackage{amsmath}
\usepackage{array}
\usepackage{xcolor}
\usepackage{arydshln}
\usepackage{siunitx}
\colorlet{tablerowcolor4}{gray!50} 

\definecolor{improveCol}{HTML}{7b3294}
\definecolor{worsenCol}{HTML}{008837}

\definecolor{entrycol}{HTML}{CD950C}
\definecolor{exitcol}{HTML}{003057}

\newcommand*{\textlabel}[2]{%
  \edef\@currentlabel{#1}
  \phantomsection
  #1\label{#2}
}

\colorlet{tableheadcolor}{gray!25} 
\colorlet{tablerowcolor}{gray!15} 
\colorlet{tablerowcolor2}{gray!45} 
\colorlet{tablerowcolor3}{gray!25} 

\newcommand{\rowcollight}{\rowcolor{LightGray}} %
\usepackage{capt-of}
\usepackage{balance,bm}
\usepackage{color, colortbl, xcolor}

\usepackage{booktabs}  
\usepackage{url}
\usepackage{hyperref}
\hypersetup{
    colorlinks=true,
}

\usepackage{colortbl}
\usepackage{arydshln}
\usepackage [english]{babel}
\usepackage [autostyle, english = american]{csquotes}
\definecolor{linkColor}{RGB}{6,125,233}
\definecolor{green}{rgb}{0.0, 0.65, 0.31}
\definecolor{bleudefrance}{rgb}{0.19, 0.55, 0.91}
\definecolor{ceruleanblue}{rgb}{0.16, 0.32, 0.75}
\definecolor{grey}{HTML}{969696}
\definecolor{violet}{HTML}{756bb1}
\definecolor{dgrey}{HTML}{01665e}
\definecolor{lgrey}{HTML}{5ab4ac}
\definecolor{dgreen}{HTML}{005a32}
\definecolor{purple}{HTML}{ae017e}

\definecolor{maskCol}{HTML}{c51b7d}
\definecolor{lrColor}{HTML}{8856a7}
\definecolor{trColor}{HTML}{d01c8b}
\definecolor{ctColor}{HTML}{4dac26}
\definecolor{brickred}{HTML}{f03b20}
\definecolor{DarkBlue}{HTML}{00008B}
\definecolor{mscolor}{HTML}{01665e}
\definecolor{nmscolor}{HTML}{bf812d}
\definecolor{lgreen}{HTML}{ccece6}
\definecolor{dolive}{HTML}{308014}

\definecolor{maskCol}{HTML}{c51b7d}
\definecolor{lrColor}{HTML}{8856a7}
\definecolor{trColor}{HTML}{d01c8b}
\definecolor{ctColor}{HTML}{4dac26}
\definecolor{brickred}{HTML}{f03b20}
\definecolor{lgreen}{HTML}{e0f3db}
\definecolor{dpink}{HTML}{CD1076}
\definecolor{pink}{HTML}{FED2D2}
\definecolor{soothinggreen}{HTML}{4dac26}

\definecolor{dblue}{HTML}{104E8B}
\definecolor{violet}{HTML}{8A2BE2}
\definecolor{mscolor}{HTML}{01665e}
\definecolor{nmscolor}{HTML}{d8b365}
\definecolor{deepgrey}{HTML}{525252}
\definecolor{dslate}{HTML}{2F4F4F}
\definecolor{dolive}{HTML}{556B2F}
\definecolor{teal}{HTML}{388E8E}
\definecolor{mscolor}{HTML}{01665e}
\definecolor{nmscolor}{HTML}{d8b365}

\definecolor{aicolor}{HTML}{018571}
\definecolor{occolor}{HTML}{ff7799}

\definecolor{srcolor}{HTML}{e34a33}
\definecolor{smcolor}{HTML}{253494}
\definecolor{srsmcolor}{HTML}{7fcdbb}
\definecolor{bothcolor}{HTML}{fe9929}
\definecolor{onecolor}{HTML}{018571}
\definecolor{marroon}{HTML}{881c1c}

\definecolor{kriyablue}{RGB}{79,126,207}
\definecolor{kriyagold}{RGB}{217,164,65}
\definecolor{kriyagreen}{RGB}{79,175,123}

\usepackage{mathtools}

\usepackage{amsmath}
\usepackage{array}
\usepackage{xcolor}
\usepackage{arydshln}
\usepackage{siunitx}
\colorlet{tablerowcolor4}{gray!50} 

\usepackage{tcolorbox}

\usepackage{xcolor}
\colorlet{tableheadcolor}{gray!25} 
\colorlet{tablerowcolor}{gray!15} 
\colorlet{tablerowcolor2}{gray!45} 
\colorlet{tablerowcolor3}{gray!25} 

\newif{\ifhidecomments}
  \hidecommentsfalse 
\ifhidecomments
    \newcommand{\jina}[1]{}
    \newcommand{\koustuv}[1]{}
    \newcommand{\connie}[1]{}
\else
    \newcommand{\jina}[1]{\textbf{\small\sffamily{\textcolor{DarkBlue}{[#1 -- Jina]}}}}
    \newcommand{\connie}[1]{\textbf{\small\sffamily{\textcolor{marroon}{[#1 -- Connie]}}}}
    \newcommand{\koustuv}[1]{\textbf{\small\sffamily{\textcolor{violet}{[#1 -- Koustuv]}}}}
  \fi

\renewcommand{\textrightarrow}{$\rightarrow$}

\colorlet{tableheadcolor}{gray!25} 
\colorlet{tablerowcolor}{gray!5} 
\definecolor{shaneColor}{HTML}{1F77B4}
\definecolor{neutralCol}{HTML}{dd1c77}
\definecolor{neutralGreen}{HTML}{31a354}
\definecolor{NewBlue}{HTML}{1879ba}
\definecolor{bleudefrance}{rgb}{0.19, 0.55, 0.91}  
\definecolor{AfTrColor}{HTML}{0868ac}  
\definecolor{BfTrColor}{HTML}{a8ddb5}  

\definecolor{AfCtColor}{HTML}{b10026}  
\definecolor{BfCtColor}{HTML}{fd8d3c}

\graphicspath{ {figures/} }

\AtBeginDocument{%
  \providecommand\BibTeX{{%
    \normalfont B\kern-0.5em{\scshape i\kern-0.25em b}\kern-0.8em\TeX}}}

\begin{document}

\title[Early Signals of Problematic AI Use: The Longitudinal Roles of Motivation and Emotional Engagement]{Early Signals of Problematic AI Use: The Longitudinal Roles of Motivation and Emotional Engagement}

\author{Jina Suh}
\affiliation{
\institution{CoralMind Institute, University of Washington}
\city{Seattle}
  \state{Washington}
  \country{USA}}
\email{jinasuh@cs.washington.edu}
\orcid{0000-0002-7646-5563}

\author{Constance Hadley}
\affiliation{
\institution{Institute for Life at Work}
\city{Belmont}
  \state{Massachusetts}
  \country{USA}}
\email{cnhadley@institutelifework.org}
\orcid{0000-0002-7646-5563}

\author{Koustuv Saha}
\affiliation{
\institution{University of Illinois Urbana-Champaign}
\city{Urbana}
  \state{Illinois}
  \country{USA}}
\email{ksaha2@illinois.edu}
\orcid{0000-0002-7646-5563}

\renewcommand{\shortauthors}{Suh et al.}

\begin{abstract}
Longitudinal evidence about the psychological impact of conversational AI use remains scarce. We report a survey study of 114 adults who reported their level of problematic AI use one year after reporting their motivations (for escape, social, entertainment, and instrumental reasons) for engaging with AI. We found high levels of AI motivation of all types predicted later problematic AI use, especially when participants turned to AI for managing their feelings and overcoming difficulties. Our findings imply that AI motivation could be used as an early warning signal of potential problematic use, especially if individuals engage with AI for emotional purposes over time. 
\end{abstract}



\maketitle

\section{Introduction}
People increasingly turn to AI systems for emotional support, self-disclosure, and companionship~\cite{zao2026hbr,skjuve2021my}, which are practices historically reserved for human relationships.
Recent survey works find that roughly one in five U.S.\ adolescents and young adults in 2025 have used AI chatbots for mental health advice~\cite{mcbain2026ai}, over a quarter of adults have used them as personal therapists~\cite{Cross2024-qy}, and more than a third of psychologists report that their patients are using AI as an additional mental health provider~\cite{abrams2026discussing}.
Therapy and companionship have become prominent use cases for general-purpose AI assistants~\cite{zao2026hbr}, and these affective conversations are no longer confined to AI systems explicitly designed as AI companions~\cite{phang2025investigating}.
On one hand, early evidence suggests that such AI systems can provide perceived empathy, reduce stress, reduce momentary loneliness, and offer accessible support~\cite{de2026ai,meng2021emotional}.
On the other hand, the very features that make them effective in these roles (e.g., responsiveness, personalization, and non-judgmental interaction) may also encourage over-reliance and unhealthy psychological dependence~\cite{xie2023friend,laestadius2024too,zhang2025dark}.

Despite growing public and scientific concern, our understanding of the long-term psychological consequences of AI-mediated emotional engagement remains limited.
Existing studies are primarily cross-sectional or short-term, offering only snapshots of user behavior and perception~\cite{yuan2026mental,skjuve2023longitudinal}. 
These studies often emphasize usage frequency or duration as primary risk factors, yet emerging evidence suggests that why individuals engage with AI---particularly for social connection or emotional escape---may be more consequential than how much they use it. 
This gap is particularly important given the scale and normalization of emotional AI use. 
As AI systems become integrated into everyday coping practices, reliance on them may develop gradually and remain under-recognized, especially against individual and cultural normalization of AI use over time.
Understanding these dynamics requires moving beyond short-term exposure effects to examine how early motivations, usage purposes, and psychological outcomes unfold over extended periods.
Few studies have followed the same people across enough time for a pattern of problematic use to develop, and fewer still have measured what preceded it.

This paper is particularly concerned with \textit{problematic AI use}, defined as a pattern of engagement with conversational AI that a person finds difficult to regulate and that interferes with other parts of life.
This construct is not entirely new. It comes from a lineage of behavioral-addiction measures, such as problematic internet use, problematic smartphone use, and internet gaming disorder, whose criteria were themselves adapted from diagnostic frameworks for substance and behavioral disorders~\cite{griffiths2005components,american2022diagnostic}.
These frameworks describe recognizable set of features that indicate that a person's engagement with a system persists past the point they intend and starts to have negative consequences in their life. 
These features include preoccupation with the activity; withdrawal-like discomfort or distress when it is unavailable; tolerance for needing more of it to obtain the same satisfaction; loss of control over time spent; decline in social, occupational, or academic functioning; loss of interest in previously enjoyable activities; and use of the activity to escape or relieve a dysphoric mood.
\citeauthor{yu2024development} adapted this structure to conversational AI in the Problematic ChatGPT Use Scale, which we use in this study. In their validation work, scores on the scale were associated with depression and psychological distress, providing evidence that the construct captures something beyond high usage.

In understanding problematic AI use, our work particularly examines \textit{AI use motivation}, which we refer to as the goals a person brings to the AI system, as a potential risk marker for problematic AI use.
Across technology addiction studies, past research has found that an individual's motivation can act as drivers of excessive or compulsive use and the development of addictive behaviors~\cite{brand2019interaction,huang2024ai}.
We hypothesize that, if problematic use is partly constituted by turning to a technology for self-regulation, then what a person is seeking from that technology, measured before any problem appears, becomes a candidate early signal.
To measure motivation, we follow \citeauthor{huang2024ai} and their four theoretically derived motivational systems for AI. \textit{Instrumental} motivation describes using AI to find information, learn, or improve one's abilities, \textit{entertainment} motivation describes using it for enjoyment or diversion, \textit{social} motivation describes using it to relieve loneliness or to have something to interact with, and \textit{escape} motivation describes using it to avoid stress, responsibility, or other problems~\cite{huang2024ai}.

Motivation, however, is not the same as behavior. Our study captures AI usage behaviors in two ways: use purpose and usage volume. 
\textit{Use purpose} refers to what the technology is actually used for, which may not necessarily match the individual's initial motivation. 
We distinguish emotional use purpose (i.e., turning to AI to manage feelings or work through difficulties) from functional use purpose (i.e., using it to accomplish tasks, learn, or work more quickly).
We measure technology usage volume through \textit{usage frequency} (i.e., how often they use the technology). 
Both use purpose and usage volume can be inferred from logs, making them accessible potential markers for downstream impacts by operators of deployed AI systems. 

Putting them together, our study examines the development of problematic AI use over a one-year span to understand the influence of motivation measured a year earlier on problematic AI use and whether the association between motivation and problematic use is operated through how much people use AI or what they use it for. 
We present our findings from a one-year follow-up study of 114 adults who had participated in a five-week experiment on social and emotional AI use in 2024~\cite{chandra2025longitudinal} with motivation measured after the experiment and again with problematic AI use a year later. 
We find that individual characteristics explain little of who develops problematic AI use. Of age, gender, race, education, household size, mental-health status, baseline AI usage, loneliness, and trait impulsivity, only age is reliably associated with it.
Motivation measured a year earlier does predict problematic AI use, and this association operates primarily through what AI is used for: emotional use purpose accounts for a substantially larger share of the effect than usage frequency for every motivation type.
This finding holds even for instrumental motivation, where the association is almost entirely accounted for by emotional use purpose; what appears to matter is whether use drifts toward emotional purposes.

In summary, this study makes the following contributions: (1) one-year prospective evidence that motivation for AI use can be used to forecast problematic AI use, and (2) evidence that this association operates through emotional use purpose rather than usage volume, including for users whose stated motivation is purely instrumental.
Our study highlights where to look for the risk of problematic AI use. 
Instead of screening users for who they are or how much they use the system, our study implies that motivation can provide an early signal and that the drift toward emotional use purpose can be a key indicator. 
These findings can be used to inform the design of systems, especially in regard to how systems motivate and reward certain AI uses.

\section{Background and Related Work}

\subsection{Social and Emotional Use of Conversational AI}

Conversational AI is increasingly used beyond informational and task-oriented purposes.
People turn to these systems to discuss personal experiences, seek emotional support, and find companionship.
A recent analysis of publicly reported AI use cases identify therapy and companionship among the prominent uses of generative AI~\cite{zao2026hbr}.
Although affective conversations appear to be concentrated among more engaged users~\cite{phang2025investigating}, these findings show that social and emotional use is no longer limited to applications explicitly designed as AI companions.
HCI research has long examined how people respond socially to computers and develop relationships with interactive systems~\cite{nass2000machines,bickmore2005establishing}.
In the context of conversational AI, prior work has shown that people disclose personal information to chatbots~\cite{ho2018psychological,lee2020hear,yoo2026ai}, perceive emotional and informational support from them~\cite{meng2021emotional,ta2020user}, and describe AI companions as friends, confidants, or relational partners~\cite{skjuve2021my,bae2021social}.
Further, AI-generated responses to mental health queries can be more structured and readable, but less grounded in lived experience than human responses~\cite{saha2026linguistic}.

Recent experimental work further suggests that interacting with an AI companion can temporarily reduce loneliness, particularly when users feel heard by the system~\cite{de2026ai}, and reduce anxiety and improve well-being in structured mental health interventions, especially among users reporting greater loneliness or lower perceived social support~\cite{shoshani2026attachment}.
\citeauthor{zhang2026interaction} found that companionship-oriented use was associated with lower well-being, particularly when interactions were intensive or highly self-disclosive~\cite{zhang2026interaction}.
However,~\citeauthor{yuan2026mental} identified a paradox in longer-term AI companion use: users often turned to these AI companions because they felt lonely, yet continued use could deepen their disconnection from offline social relationships and reinforce the loneliness that initially motivated their use~\cite{yuan2026mental}.
Together, this body of work demonstrates how conversational AI can provide accessible and responsive support when human interaction is unavailable or less preferred.

Further, prior research has raised concerns about how such interactions develop over time.
Continued interaction can deepen self-disclosure and perceived relational closeness~\cite{skjuve2023longitudinal}, while disruptions to an established AI companion can produce feelings of distress, rejection, or loss~\cite{laestadius2024too}.
Evidence from clinically oriented conversational agents indicates reductions in depression and psychological distress, although their long-term effects remain uncertain~\cite{li2023systematic}.
Prior work also noted how engagement, personification, trust, and attachment may contribute to psychological dependence on an AI companion~\cite{xie2023friend}.
Building on this body of work, this study examines how motivations for social and emotional conversational AI use relate to problematic use of conversational AI systems.

\subsection{Problematic Use in Human--AI Interaction}

Much of prior work in human--AI interaction has focused on the concept of \textit{reliance}.
\citeauthor{lee2004trust} described AI reliance as a behavioral response to AI recommendations, distinct from trust as an attitude toward the system~\cite{lee2004trust}.
Appropriate reliance involves accepting correct recommendations while recognizing when the system is wrong; however, people frequently follow incorrect recommendations when independently evaluating them requires additional cognitive effort~\cite{parasuraman1997humans,buccinca2021trust}.
Providing explanations may improve users' understanding of an AI recommendation, but does not consistently reduce over-reliance~\cite{bansal2021does,vasconcelos2023explanations}.
Human-centered AI research therefore emphasizes preserving meaningful human agency and control~\cite{shneiderman2020human}.
Related work further suggests that reliance develops through continued interaction, as systems communicate their capabilities and limitations, support correction and recovery, and users revise their assumptions about AI within broader social and organizational contexts~\cite{amershi2019guidelines,wang2021towards,ehsan2023charting}.
While reliance concerns whether a person appropriately accepts or rejects a system's output on a given task, \textit{problematic use}---our primary focus---concerns whether a person can regulate their engagement with the system at all.

Research on addiction, particularly computing technology-based forms, can inform our understanding of problematic AI use.
Clinical definitions of substance-use disorders emphasize impaired control, unsuccessful attempts to reduce use, tolerance, withdrawal, and continued use despite social, occupational, or health consequences~\cite{american2022diagnostic}
Internet gaming disorder criteria and subsequent problematic internet, social media, and smartphone-use scales share a common structure of preoccupation, withdrawal, tolerance, loss of control, functional impairment, displacement of other interests, and mood-regulatory use~\cite{american2022diagnostic,griffiths2005components}.
Prior to ChatGPT,~\citeauthor{kardefelt2017conceptualizing} cautioned against treating frequent technology use as problematic in the absence of impaired control or disrupted functioning.
This caution applies with particular force to conversational AI, where heavy use may remain purposeful and controlled.

Extending these ideas to conversational AI,~\citeauthor{yu2024development} developed the Problematic ChatGPT Use Scale to capture maladaptive engagement, including preoccupation with ChatGPT and difficulty regulating its use, interference with tasks and sleep, concealment of use from others, and turning to the system to relieve helplessness or anxiety~\cite{yu2024development}.
In their validation sample, scores on that scale were associated with depression and psychological distress, providing evidence that the scale captures more than usage frequency.
Recent critiques nevertheless argue that intensive ChatGPT use should not be pathologized without clear evidence of impaired control, psychological distress, negative consequences, and functional impairment~\cite{ciudad2025people}.

Recent studies have begun to examine how problematic engagement develops around conversational AI specifically.
\citeauthor{hu2023social} found that social anxiety was associated with problematic conversational AI use through loneliness and rumination~\cite{hu2023social}.
\citeauthor{shen2025dark} identified interface patterns that can encourage prolonged or compulsive chatbot engagement~\cite{shen2025dark}.
In contrast,~\citeauthor{chandra2025longitudinal} observed increases in attachment and perceived empathy after five weeks of social and emotional AI use, but no corresponding increase in dependence~\cite{chandra2025longitudinal}.
Building on this body of work, we examine whether motivations measured up to a year earlier predict problematic AI use, and whether that association operates through emotional use purpose or through usage volume.

\subsection{Motivational and Longitudinal Change in AI Use}

Users' experiences with interactive systems can change as technologies become integrated into everyday life.
\citeauthor{karapanos2009user} distinguished between orientation, incorporation, and identification, showing how users move from initial learning and exploration toward more stable and personally meaningful forms of technology use~\cite{karapanos2009user}.
\citeauthor{kujala2011ux} similarly demonstrated that users' evaluations change over time and are shaped by meaningful positive and negative experiences rather than initial impressions alone~\cite{kujala2011ux}.
More recently,~\citeauthor{long2025facilitating} argued that longitudinal evaluation is particularly important for AI systems because people learn, adapt, and repurpose these systems through continued use~\cite{long2025facilitating}.
These studies show why early evaluations may not reflect how people eventually incorporate AI into their routines.

Longitudinal studies of AI use have examined these changes across different contexts.
For instance, re-engaging participants from an earlier study one year later,~\citeauthor{kobiella2025efficiency} examined how knowledge workers moved from initial experimentation with LLMs toward more selective forms of integration shaped by efficiency, control, and fulfillment~\cite{kobiella2025efficiency}.
In a year-long study of AI use among cancer specialists,~\citeauthor{ehsan2026future} found that immediate improvements in performance could conceal gradual reductions in expertise and independent judgment~\cite{ehsan2026future}.
These findings illustrate how the longer-term consequences of AI use may differ from the benefits or concerns users report during early adoption.

Prior work has examined why people turn to conversational AI.
\citeauthor{skjuve2024people} identified productivity, novelty, creative work, learning, entertainment, and social interaction and support as distinct motivations for using ChatGPT~\cite{skjuve2024people}.
Moving beyond general motivations,~\citeauthor{huang2024ai} distinguished between instrumental, entertainment, social, and escape motivations for AI use~\cite{huang2024ai}.
These findings suggest that the longer-term implications of AI use may depend on why people use these systems, and not only on how frequently they use them.
Building on this work, we examine how earlier motivations are associated with problematic AI use one year later.

\section{Methods}
The present study is a one-year follow-up of a five-week experimental study conducted in 2024 by~\citet{chandra2025longitudinal}.
We first briefly summarize the design of that prior study and then describe the aims, measures, retained sample, and analysis methods for the present study.

\subsection{Study Design}\label{sec:methods:study-design}

\subsubsection{Prior Study}\label{sec:methods:study-background}

In the 2024 experiment, participants in the Baseline Usage (BU) arm ($n$=60) used the internet and AI as they normally would. Participants in the Active Usage (AU) arm ($n$=89) were assigned one of four platforms (ChatGPT Plus, Copilot Pro, Gemini Advanced, or Pi) and asked to use it for at least 10 minutes daily for five weeks across five social and emotional scenarios, such as talking through personal struggles, managing stress, and building emotional connection with the AI. 
This experiment protocol lasted five weeks between July and August of 2024.
Although the AU arm reported increased attachment, perceived empathy, and satisfaction, neither arm showed a measurable increase in self-reported dependency over the five weeks, which motivated a longer-term follow-up. 
Full details of the 2024 experiment can be found in prior work~\cite{chandra2025longitudinal}, including information about the protocol, cohort, between-subjects assignment, and findings.

\subsubsection{Present Study}\label{sec:methods:present-study}
The present study is a one-year follow-up that focuses on the predictors of problematic AI use.
The follow-up was a single online Qualtrics survey administered to the ~\cite{chandra2025longitudinal} study participants who consented to be recontacted.
After obtaining permission from the prior study's research team~\cite{chandra2025longitudinal} and approval from our IRB for the follow-up protocol, including participant recontact and data linkage, we invited participants from the prior study who had previously consented to be recontacted. 
Before beginning the follow-up survey, participants received a description of the new study and provided separate informed consent to participate and to link their follow-up responses to their prior responses. 

The full study, therefore, spans three measurement waves in approximately a 14-month period. Time 0 (T0; June, 2024) is the pre-experiment screening and intake. Time 1 (T1; August, 2024) is the post-experiment exit survey approximately 5 weeks after T0. Time 2 (T2; August, 2025) is the one-year follow-up approximately twelve months after T1.
At follow-up, we retained the original arm assignments without re-randomization, resulting in the 2025 split being 69 AU and 45 BU participants (see Section~\ref{sec:methods:participants}).
The prior experiment was exploratory and focused on measuring the effects of five weeks of assigned social and emotional AI use on their experiences and perceptions. 
Our current study asks: given the same people, what predicts problematic AI use a year later?
Therefore, the primary outcome of the present study is problematic AI use~\cite{yu2024development} measured at T2. 
We ask three questions:

\begin{itemize}
\item \textbf{RQ1.} Do individual characteristics explain who develops problematic AI use?
\item \textbf{RQ2.} Does the level and/or type of motivation for AI use, measured a year earlier, predict problematic AI use?
\item \textbf{RQ3.} Does \textit{how much} people use AI and/or \textit{what} they use it for explain the relationship between motivation and problematic use?
\end{itemize}

After the consent section, the survey flow was roughly organized as follows: general attitudes toward AI, AI dependence, liking of AI, a platform-by-platform usage frequency matrix, AI use purpose, problematic AI use, social connectedness, loneliness, and perceived negative impacts of AI, with short open-ended prompts throughout. 
The survey re-administered the measures from the original study with several additions and included measures we do not analyze in this paper. 
The measures used in this paper and their psychometric properties are described in Section~\ref{sec:methods:measures}, and the rest are described in Appendix~\ref{appx:methods:unanalyzed}.
Regardless of the original assignment, all participants received the same survey and were compensated US\$20 for survey completion. The survey was estimated to take 45 minutes, and the median completion time was 19 minutes.  

Both the original study and the follow-up were reviewed and approved by the lead institute's Ethics Review Board. The study design and exploratory research questions were preregistered on the Open Science Framework prior to data collection.

\subsection{Measures}\label{sec:methods:measures}

Table~\ref{tab:measures} summarizes the measures used in this study, including the wave of measurement, number of items, response scale, and reliability statistics.
Full item text, response options, scoring rules, and additional statistics can be found in Appendix~\ref{appx:methods:measures}.
Internal consistency is reported using Cronbach's $\alpha$ for scales with three or more items and using the Spearman--Brown coefficient for the two-item scale. 

\begin{table}[htbp]
\centering
\small
\sffamily
\caption{Measures used in this paper, including the waves the measure was collected, number of items in the measure, scoring scale, reliability (Cronbach's $\alpha$, or the Spearman--Brown coefficient), mean ($M$) and standard deviation ($SD$) at the wave used in the primary analysis (T1 for motivation, T2 otherwise), and the role the measure plays in the analysis. $^{a}$Usage frequency uses a different instrument at T0/T1 (5 platforms, daily scale) than at T2 (13 platforms, weekly scale); the two are not on a common metric.}
\label{tab:measures}
\begin{tabular}{llccccl}
\textbf{Construct} & \textbf{Wave(s)} & \textbf{Items} & \textbf{Scale} & \textbf{Rel.} & \textbf{$M$ ($SD$)} & \textbf{Role} \\
\hline
\rowcollight \multicolumn{7}{l}{\textit{Primary outcome}} \\
\quad Problematic AI use & T2 & 10 & 1--5 & .87 & 1.62 (0.68) & Outcome \\
\hdashline
\rowcollight \multicolumn{7}{l}{\textit{Focal predictors}} \\
\quad Motivation: escape & T0, T1, T2 & 3 & 1--4 & .94 & 1.70 (0.93) & Predictor \\
\quad Motivation: social & T0, T1, T2 & 3 & 1--4 & .95 & 1.79 (0.91) & Predictor \\
\quad Motivation: instrumental & T0, T1, T2 & 3 & 1--4 & .82 & 3.44 (0.65) & Predictor \\
\quad Motivation: entertainment & T0, T1, T2 & 3 & 1--4 & .89 & 2.50 (0.89) & Predictor \\
\quad Motivation: overall & T0, T1, T2 & 12 & 1--4 & .91 & 2.36 (0.65) & Predictor \\
\hdashline
\rowcollight \multicolumn{7}{l}{\textit{Candidate mediators}} \\
\quad Use purpose: emotional & T2 & 2 & 1--5 & .77 & 2.68 (1.23) & Mediator \\
\quad Use purpose: functional & T2 & 6 & 1--5 & .90 & 3.97 (0.96) & Mediator \\
\quad Use purpose: external & T2 & 5 & 1--5 & .65 & 1.68 (0.66) & Mediator (control) \\
\quad Usage frequency$^{a}$ & T0, T1, T2 & 1 & 1--8 & --- & 5.96 (1.85) & Mediator \\
\hdashline
\rowcollight \multicolumn{7}{l}{\textit{Covariates}} \\
\quad Age & T2 & 1 & 1--6 & --- & --- & Covariate \\
\quad Education & T2 & 1 & 1--4 & --- & --- & Covariate \\
\quad Mental-health condition & T0 & 1 & 0/1 & --- & --- & Covariate \\
\hdashline
\rowcollight \multicolumn{7}{l}{\textit{Secondary (supplement)}} \\
\quad AI dependence & T0, T1, T2 & 5 & 5--20 & --- & --- & Secondary outcome \\
\quad Attitudes (positive, negative) & T0, T1, T2 & 20 & 1--5 & --- & --- & Arm contrast \\
\quad Social connectedness & T2 & 8 & 1--5 & --- & --- & Arm contrast \\
\quad Loneliness & T0, T2 & 3 & 3--9 & --- & --- & Arm contrast \\
\hline
\end{tabular}
\end{table}

\subsubsection{Primary outcome}
\textit{Problematic AI use} (T2 only) was measured with 10 items adapted from~\citeauthor{yu2024development}'s Problematic ChatGPT Use Scale~\cite{yu2024development}, rated on a 5-point agreement scale and averaged (range 1--5; $\alpha$=.87).
We depart from the source scale in four ways. We administered nine of the original 11 items, omitting the `anxiety and irritability when unable to access' item and the `attempted to reduce usage without success' because each overlaps conceptually with an AI dependence item also administered in this survey (withdrawal and loss of control). We added one item (``I become fully absorbed when using AI conversational agents''). 
We used a 5-point rather than 4-point agreement scale, and a mean rather than summed score, to better match other measures in the survey.
Because of the rescaling and mean scoring, our scores are not directly comparable to the summed 11-item distribution reported by~\citeauthor{yu2024development} or to any published cutoff.
A sum-scored sensitivity analysis is reported in Appendix~\ref{appx:methods:sumscore}.

\subsubsection{Motivation for AI use}
We measured \textit{motivation for AI use} at all three waves using~\citeauthor{huang2024ai}'s scale without alteration. Each of the four subscales (escape, social, instrumental, entertainment) includes three items each, scored on a 4-point agreement scale and averaged per subscale (range 1--4). 
We also computed an \textit{overall motivation} composite as the mean of the four subscale means, and this composite score was not part of the original scale.
Reliability was high at every wave (T1: $\alpha$=.94 escape, .95 social, .82 instrumental, .89 entertainment).
A confirmatory factor analysis of the four-factor structure fit well ($\chi^{2}/df$=1.64, CFI=.97, RMSEA=.075, SRMR=.059; standardized loadings .72--.97).

\subsubsection{AI use purpose}
\textit{AI use purpose} (T2 only) captures what participants use AI for, with the stem ``I use AI conversational agents\ldots'' rated on a 5-point agreement scale and averaged per subscale (range 1--5). The items are motivated by and adapted from~\citeauthor{wolf2024chatgpt}. We captured three types of AI use purposes.
\textit{Emotional purpose} comprises two items: ``to overcome certain difficulties'' and ``to manage my feelings'' (Spearman--Brown=.77).
\textit{Functional purpose} comprises six items covering task completion and learning (e.g., ``because it helps me to accomplish tasks more quickly,'' ``to enhance my learning and knowledge''; $\alpha$=.90).
\textit{External purpose} comprises five items reflecting social pressure to use AI (e.g., ``because many people I know use it;'' $\alpha$=.65).

\subsubsection{Usage amount}
\textit{Usage frequency} at T2 is the maximum self-reported frequency across 13 AI platforms, spanning general-purpose assistants and companion-oriented apps (Appendix~\ref{appx:methods:platforms}), each rated on an 8-point scale (1=never to 8=multiple times per day).
At T0 and T1, usage was measured with a 5-platform matrix on a daily frequency scale (1--5). 
At every wave the analysis variable is the maximum frequency a participant reported for any single platform to capture the intensity of use.
Because the two instruments differ in platform coverage and response metric, T0/T1 and T2 usage values are not on a common scale and we make no cross-wave usage comparisons.

\subsubsection{Individual characteristics and covariates}

Age was collected as six age buckets (e.g., 18-25, 26-35, 56-65, 66+) and treated as continuous ordinal.
Education was collapsed into four groups: high school or less, some college or associate's degree, bachelor's degree, and graduate degree. 
Race and ethnicity were collected as a multi-select item and was then transformed into a binary variable (White vs non-White) given the skewed data.
Age and education were two variables re-collected at T2. All other variables were collected at T0 from the original study.
Table~\ref{tab:sample} summarizes the participant characteristics.
We included age, education, and baseline mental-health condition, as covariates for our models.
We chose as covariates age and education because they were two demographic characteristics the original study did not balance on, and mental health condition is included for its potential relevance to the outcome. 
These covariates were fixed a priori.
We deliberately left out baseline AI usage despite its correlation with the outcome ($r$=.22) because it is a competing variable that our \textbf{RQ3} analysis is trying to examine directly.
Nonetheless, to confirm that our choice of covariates did not alter our results, we conducted a sensitivity analysis under different specifications of covariates. These alternate specifications included having no covariates, adding baseline AI usage to the selected three, and using a set of variables found to be significantly correlated with the outcome. We found these alternate specifications to not change our results (Appendix~\ref{appx:methods:controls}).
The remaining characteristics, such as gender, race, household size, baseline AI usage, loneliness, and trait impulsivity, entered only the \textbf{RQ1} analysis of whether individual characteristics predict problematic AI use .

\subsection{Participants}\label{sec:methods:participants}

\begin{table}[htbp]
\centering
\small
\sffamily
\caption{Participant characteristics from the one-year follow-up study ($N$=114). Age and education were collected at T2; all other variables were collected at T0.}
\label{tab:sample}
\begin{tabular}{lc}
\textbf{Characteristic} & \textbf{$n$ (\%)} \\
\hline
\rowcollight \multicolumn{2}{l}{\textit{Gender}} \\
\quad Male & 54 (47.4\%) \\
\quad Female & 60 (52.6\%) \\
\rowcollight \multicolumn{2}{l}{\textit{Age}} \\
\quad 18--25 & 3 (2.6\%) \\
\quad 26--35 & 35 (30.7\%) \\
\quad 36--45 & 39 (34.2\%) \\
\quad 46--55 & 23 (20.2\%) \\
\quad 56--65 & 10 (8.8\%) \\
\quad 66+ & 4 (3.5\%) \\
\hdashline
\rowcollight \multicolumn{2}{l}{\textit{Race and ethnicity}} \\
\quad White & 84 (73.7\%) \\
\quad Black or African American & 11 (9.6\%) \\
\quad Asian & 6 (5.3\%) \\
\quad Hispanic or Latino & 3 (2.6\%) \\
\quad Multiracial or another identity & 10 (8.8\%) \\
\hdashline
\rowcollight \multicolumn{2}{l}{\textit{Education}} \\
\quad High school or less & 15 (13.2\%) \\
\quad Some college or associate's degree & 44 (38.6\%) \\
\quad Bachelor's degree & 34 (29.8\%) \\
\quad Graduate degree & 21 (18.4\%) \\
\hdashline
\rowcollight \multicolumn{2}{l}{\textit{Household income}} \\
\quad Less than \$30,000 & 26 (22.8\%) \\
\quad \$30,000--\$59,999 & 36 (31.6\%) \\
\quad \$60,000--\$99,999 & 25 (21.9\%) \\
\quad \$100,000 or more & 27 (23.7\%) \\
\hdashline
\rowcollight \multicolumn{2}{l}{\textit{Mental health condition}} \\
\quad No & 60 (52.6\%) \\
\quad Yes & 54 (47.4\%) \\
\hline
\end{tabular}
\end{table}

Eligible participant pool comes from 149 participants who completed the 5-week study.
The 2024 study cohort was recruited from the U.S. Prolific panel based on a stratified sampling to balance gender identity, existence of any mental health condition, and prior AI usage tendency. 
Of the 149 eligible participants, 114 completed the same one-year follow-up survey (76.5\% retention).
Table~\ref{tab:sample} summarizes the retained sample's characteristics.
The retained sample was balanced on gender (52.6\% female; 47.4\% male).
Almost half (47.4\%) reported a mental-health condition at screening in 2024, rising to 67.5\% at follow-up (see Section~\ref{sec:limitations}).
The sample skewed toward ages 26--45 (64.9\%) and was predominantly White (73.7\%).

To assess potential selection bias, we compared 114 completers with the 35 non-completers on baseline characteristics.
For between-group comparisons, we used Welch two-sample t-tests with Cohen's d as the effect size, adjusting for multiple comparisons using Benjamini--Hochberg-correction.
We found that completers were older ($d$=0.80), less impulsive ($d$=$-$0.55), and less likely to report a mental-health condition (47\% vs.\ 74\%) at T1; all $q$<.05).
Baseline AI usage was lower among completers but did not survive correction ($q$=.108).
Because problematic AI use was administered only at T2, we cannot compare the two groups on the primary outcome, so we use the proxy AI dependence measure at T1. 
Non-completers did not differ on any motivation subscale or on baseline AI dependence ($|d|\leq0.22$).

In the preliminary analysis of the retained cohort, we asked whether the original experimental arms had any lasting effect on the follow-up outcomes. 
Across all preregistered confirmatory contrasts between experiment arms (Welch t-tests at T2 on positive and negative AI attitudes, AI dependence, problematic use, social and escape motivation, social connectedness, and loneliness), we found none to be significant after Benjamini-Hochberg correction (all $q \geq .20$, $|d| \leq 0.43$; Appendix~\ref{appx:methods:arms}).
Loneliness showed the only nominal difference (unadjusted $p = .024$) in that the AU group reported higher loneliness.
Because the five-week experimental nudge left no detectable difference a year later, we pooled the two assignment groups in the subsequent analyses, including assignment as covariates in sensitivity analyses.

\subsection{Analysis}

We organize our analysis methods by research questions.

For \textbf{RQ1}, we used a regression model with the full set of individual characteristics (age, gender, race, education, household size, mental-health status, baseline AI usage, loneliness, and trait impulsivity) as predictors. 

For \textbf{RQ2}, we modeled problematic AI use on each motivation measure using ordinary least squares regression, controlling for age, education, and baseline mental-health status:
\begin{equation*}
    \text{ProblematicUse}_{T2} \sim \text{Predictor}_{w} + \text{Age} + \text{Education} + \text{MentalHealth}_{T0},
\end{equation*}
where the predictor is a motivation measure at wave $w \in \{T1, T2\}$, usage frequency, or an individual characteristic; the same models at T0 are reported in Appendix~\ref{appx:methods:t0}.
Continuous predictors were standardized so coefficients are comparable across models, and we report adjusted $R^{2}$ and 95\% confidence intervals throughout.
We explored variants of this model for additional checks. For these, we added terms without changing the modelling approach: additive models entered a motivation measure and usage frequency together; joint models entered all four motivation subscales simultaneously; and moderation models added a motivation~$\times$~usage interaction.
We report variance inflation factors for every model with more than one substantive predictor.
Where a result could depend on the outcome's distribution or on individual observations, we re-analyzed it using rank order rather than raw scores.

For \textbf{RQ3}, we estimated mediation models with 95\% percentile bootstrap confidence intervals on the indirect effects (1000 resamples with fixed seed).
We ran single-mediator models that routed each T1 motivation measure through each of four T2 mediator candidates: emotional, functional, and external use purpose, and usage frequency.
We also ran a joint model that entered all four mediators simultaneously to estimate each indirect effect holding the other three constant. 
To compare these indirect effects, we bootstrapped pairwise contrasts between the emotional-purpose indirect effect and each rival channel.
Because these mediators and the outcome are measured in the same wave, the models cannot establish temporal ordering~\cite{maxwell2007bias}. Instead, we compare across mediators to see which is most consistent with the data.

Throughout, we controlled for false-discovery rate from multiple hypothesis testing by applying the Benjamini--Hochberg correction within analysis families. Corrected $p$ values are indicated as $q$. 
For correlations, we use Pearson's $r$. 
All quantitative analyses were conducted in Python (pandas, numpy, scipy, statsmodels, semopy). 
AI-assisted language editing was used to improve grammar and readability. AI was used to assist with the formatting of tables and figures and to generate accessibility notes. The study design, analyses, interpretations, and experiments were conducted fully by the authors. 
\begin{figure}[!ht]
\centering
\includegraphics[width=0.8\linewidth]{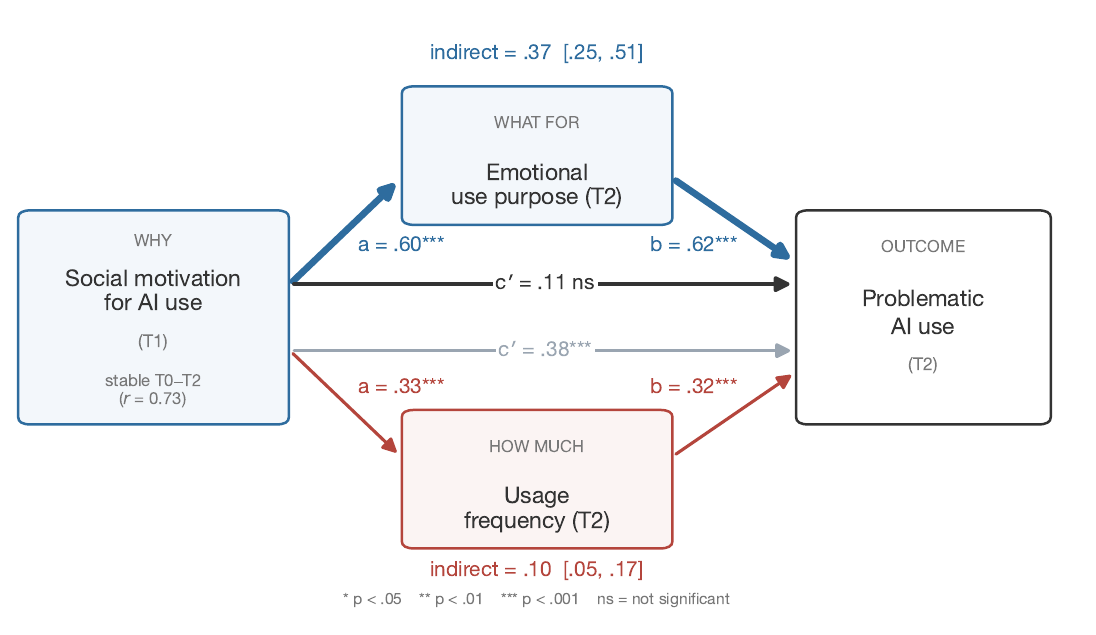}
\caption{Path diagram of the main result, with social type of AI motivation as a representative case for the other motivations tested (escape, entertainment, and instrumental). Figure shows standardized coefficients from two single-mediator models. Indirect effects also report 95\% bootstrapped confidence intervals ($N$=1000). The association between motivation at T1 and problematic AI use at T2 runs mainly through emotional use purpose (blue) rather than usage frequency (red). The direct path $c'$ is no longer distinguishable from zero once emotional use purpose is in the model (upper triangle) but remains significant with usage frequency as the mediator (lower triangle). * $p$<.05; ** $p$<.01; *** $p$<.001; ns = not significant.}
\label{fig:study_overview}
\end{figure}

\section{Results}

Our study was aimed at understanding how motivation to use AI predicts problematic AI use a year in advance, and what mediates the relationship between these variables. 
As a recap, we examined problematic AI use as a primary outcome because it measures a person's difficulty in regulating their use, which may be a precursor to a long-term psychological impact. 
We first examined whether the risk of problematic AI use comes from the person (\textbf{RQ1}; \autoref{sec:results:characteristics}). 
We then examined motivation as a potential predictor of problematic AI use (\textbf{RQ2}; \autoref{sec:results:motivation}). 
Lastly, we examined the role of use purpose or usage frequency (\textbf{RQ3}; \autoref{sec:results:channel}).

\subsection{Descriptive statistics}\label{sec:results:descriptives}

We first present the descriptive statistics of how the outcome is distributed and how stable the predictor is.
Problematic AI use was generally low in this sample ($M$=1.62, $SD$=0.68 on a 1--5 scale; median 1.40, skewness 1.39).
Most participants reported disagreement with the problematic-use items. 
23.7\% disagreed with every item, 78.1\% averaged below ``somewhat disagree'' (2.0), and only 6.1\% ($n$=7) had average score above the neutral midpoint (3.0). No participant averaged ``somewhat agree'' (4.0) or higher, and the observed maximum was 3.90.
Accounting for this distribution, we found that our results still hold under rank-based estimation (Appendix~\ref{appx:results:rank}). 
We also found that problematic AI use was correlated with concurrent loneliness at T2 ($r$=.29. $p$=.002), and the association held with age, education, and baseline mental-health status as covariates ($\beta$=.33, $p$<.001).
This correlation tracks the original scale validation study that found cross-sectional co-occurrence with psychological distress~\cite{yu2024development}.
Over the study period, we found that participants retained their relative positions on motivation for AI use (Pearson $r$(T0,T2)=.73; $r$(T1,T2)=.75), indicating that motivation for AI use is a relatively stable individual difference.

\subsection{\textbf{RQ1}: Individual characteristics explain little}\label{sec:results:characteristics}

We jointly modeled problematic use on age, education, gender, mental-health status, loneliness, impulsivity, race, household size, and baseline usage. 
This model only explained 14\% of adjusted variation, and only age was found to be significant after correction, where older participants reported more problematic use  ($\beta$=0.24, $q$=.001; Appendix~\ref{appx:results:characteristics}). 
Table~\ref{tab:descriptives} reports each individual characteristic's correlation with problematic AI use and other analysis variables. 
Here, age is also the only significant correlate ($r$=.31).
Other than age, no other variable predicted problematic AI use one year later or showed significant correlation with problematic AI use.
We interpret this finding as inconclusive as nine predictors with a small sample size ($n$=114) is not sufficiently powered to detect small effects.

\begin{table}[t]
\centering
\footnotesize
\sffamily
\caption{Descriptive statistics, reliability, and correlations of the study variables. Reliability is measured using Cronbach's alpha. Emotional use purpose reliability is Spearman--Brown (2 items); the single-item usage frequency measure has no internal-consistency estimate. The lower panel gives each individual characteristic's correlation with variables 1--10. Correlations are Pearson and uncorrected, with $|r| \geq .19$ is significant at $p<.05$, $|r| \geq .24$ at $p<.01$, and $|r| \geq .31$ at $p<.001$.}
\label{tab:descriptives}
\begin{tabular}{llccccccccccccc}
\multicolumn{2}{l}{\textit{\textbf{Variables}}} & \textbf{$M$} & \textbf{$SD$} &\textbf{ Rel.} & \textbf{1} & \textbf{2} & \textbf{3} & \textbf{4} & \textbf{5} & \textbf{6} & \textbf{7} & \textbf{8} & \textbf{9} & \textbf{10} \\
\hline
1 & Problematic use (T2) & 1.62 & 0.68 & .87 &  &  &  &  &  &  &  &  &  &  \\
2 & Escape motivation (T1) & 1.70 & 0.93 & .94 & .43 &  &  &  &  &  &  &  &  &  \\
3 & Social motivation (T1) & 1.79 & 0.91 & .95 & .51 & .59 &  &  &  &  &  &  &  &  \\
4 & Entertainment motivation (T1) & 2.50 & 0.89 & .89 & .42 & .52 & .50 &  &  &  &  &  &  &  \\
5 & Instrumental motivation (T1) & 3.44 & 0.65 & .82 & .31 & .20 & .31 & .54 &  &  &  &  &  &  \\
6 & Overall motivation (T1) & 2.36 & 0.65 & .91 & .55 & .79 & .81 & .83 & .61 &  &  &  &  &  \\
7 & Emotional use purpose (T2) & 2.68 & 1.23 & .77 & .69 & .41 & .60 & .37 & .36 & .57 &  &  &  &  \\
8 & Functional use purpose (T2) & 3.97 & 0.96 & .90 & .37 & .16 & .20 & .30 & .46 & .34 & .48 &  &  &  \\
9 & External use purpose (T2) & 1.68 & 0.66 & .65 & .20 & .05 & $-$.02 & .04 & .03 & .03 & .17 & .22 &  &  \\
10 & Usage frequency (T2) & 5.96 & 1.85 & --- & .44 & .20 & .28 & .20 & .30 & .31 & .47 & .59 & .07 &  \\
\hline
\rowcollight \multicolumn{2}{l}{\textit{Individual characteristics}} & $M$ & $SD$ & Rel. & 1 & 2 & 3 & 4 & 5 & 6 & 7 & 8 & 9 & 10 \\
 & Age & & & & .31 & $-$.09 & .11 & .10 & .16 & .08 & .01 & .03 & $-$.00 & .05 \\
 & Education & & & & .09 & $-$.06 & $-$.12 & $-$.13 & .01 & $-$.10 & .01 & .22 & .12 & .19 \\
 & Gender (female) & & & & .01 & $-$.07 & $-$.21 & $-$.06 & $-$.02 & $-$.12 & $-$.11 & $-$.06 & $-$.01 & $-$.10 \\
 & Mental health (T0) & & & & .11 & .05 & .16 & $-$.05 & $-$.17 & .02 & .17 & $-$.06 & $-$.08 & $-$.10 \\
 & Loneliness (T0) & & & & .03 & .02 & .16 & $-$.03 & $-$.09 & .03 & .12 & $-$.09 & $-$.09 & $-$.12 \\
 & Impulsivity (T0) & & & & .06 & .24 & .25 & $-$.00 & $-$.06 & .16 & .19 & $-$.14 & $-$.01 & $-$.06 \\
 & Race (White) & & & & $-$.18 & $-$.12 & $-$.18 & $-$.02 & .01 & $-$.11 & $-$.23 & $-$.07 & $-$.01 & $-$.26 \\
 & Household size (T0) & & & & .00 & .01 & $-$.01 & $-$.01 & $-$.02 & $-$.01 & .01 & .05 & $-$.00 & .01 \\
 & Baseline usage (T0) & & & & .22 & .19 & .24 & .26 & .15 & .28 & .08 & .15 & .00 & .32 \\
\hline
\end{tabular}
\end{table}

\subsection{\textbf{RQ2}: AI use motivation predicts problematic AI use a year later}\label{sec:results:motivation}
We examined AI use motivation across the four subscales (instrumental, entertainment, social, and escape)~\cite{huang2024ai} and the overall composite at T1 and T2.
We predicted each motivation measure's contribution to the problematic use at T2~\cite{yu2024development}, controlling age, education, and baseline mental-health status. 
We found that all five coefficients were positive (overall $\beta$=.367, social $\beta$=.328, escape $\beta$=.312, entertainment $\beta$=.287, and instrumental $\beta$=.201; all $p$<.001). 
Even though we see that the coefficients are ordered (i.e., social is largest, instrumental is smallest), we cannot rank the subscales because the subscales exhibit distributional variance that limit direct comparison of standardized coefficients: escape and social are floor-bound (51.8\% and 44.7\% of the sample with $M$=1.0 on a 1--4 scale), whereas instrumental is ceiling-bound (37.7\% at $M$=4.0) with a considerably smaller standard deviation (0.65 against 0.89--0.93 for the others).
Because the escape and social subscales are floor-bound, we also compared participants endorsing any escape or social motivation at T1 with those at the subscale floor (a mean of 1.0 across all three items). We found that participants endorsing any escape or social motivation at T1 scored about half a point higher on the 5-point problematic AI use scale ($\beta$=.519 and $\beta$=.533 respectively).

\begin{figure}[t]
\centering
\includegraphics[width=0.7\linewidth]{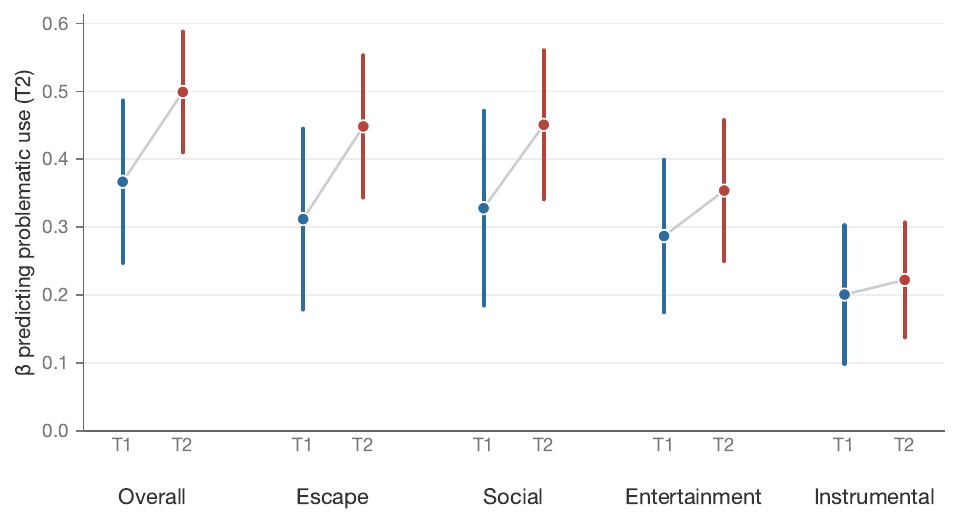}
\caption{Motivation at T1 and T2 as a predictor of problematic AI use at T2. Points are standardized coefficients from separate ordinary least squares models, with age, education, and baseline mental-health status as covariates. Vertical bars indicate 95\% confidence intervals. Prospective estimates (T1) are shown in blue; the same-wave estimate (T2) in red. The prospective association is weaker than the concurrent one but is available a year in advance.}
\label{fig:motivation-waves}
\end{figure}

We also did not find distinguishable difference between motivation subscales.
We compared all four coefficients in a joint model and found no statistically significant differences across pairwise or omnibus specifications (Appendix~\ref{appx:results:jointmodel}).
While variance inflation factors remain modest (1.51--1.98), we found no individual subscale to retain a significant coefficient (escape $\beta$=.221, social $\beta$=.247, entertainment $\beta$=.135, instrumental $\beta$=.089; all $p\geq$.07).
For all five versions of motivation, we found the measures predict problematic use better in the concurrent timeframe (T2) than a year before (T1) (adjusted $R^2$=0.635, vs 0.386 at T1 for overall motivation) as shown in Figure~\ref{fig:motivation-waves}. 

As a secondary analysis, we examined whether the T1 demographic and dispositional variables captured (age, education, gender, mental-health status, loneliness, impulsivity, race, household size, and baseline usage frequency) predicted rates of motivation in that same time period. 
In a joint model using all individual characteristic variables, we found that these individual characteristics only explain 9\% of adjusted variance in overall motivation at T1, at T2, or in its change, with none found to be significant (Appendix~\ref{appx:results:who}).
Table~\ref{tab:descriptives} describes how individual characteristics are correlated with motivation subscales. After correction, none of these characteristics remained significant.

\subsection{\textbf{RQ3}: The association runs through what AI is used for more than how much}\label{sec:results:channel}

Our next analysis examined what the motivation signal operates through, i.e., if there is an intermediary signal that carries the association between motivation and problematic use.
We tested two candidate channels: (1) how much people use AI (measured by the maximum weekly frequency across 13 platforms at T2), and (2) what they use it for (measured by three use purpose scales at T2).

First, we found that usage frequency did not absorb the signal fully. 
We found that usage frequency predicted problematic use on its own ($\beta$=0.30, adjusted $R^2$=0.28).
However, when we added usage frequency to each motivation model, we saw that only about one fifth of the motivation coefficient was removed, and every motivation coefficient remained significant after correction (Appendix~\ref{appx:results:attenuation}).
This means that, for participants with similar usage frequency, more motivated participants reported more problematic use; similarly, at equal motivation, heavier users reported more problematic use. 

When we examined motivation~$\times$~frequency interactions, we found only escape~$\times$~T2 frequency to be nominally significant after correction ($q$=.048). However, this interaction effect could not be reproduced using a rank-based estimator ($\beta$=$-$.003, $p$=.97).
Therefore, we could not find reliable evidence for this interaction effect with the current sample size.

\begin{figure}[t]
\centering
\includegraphics[width=0.8\linewidth]{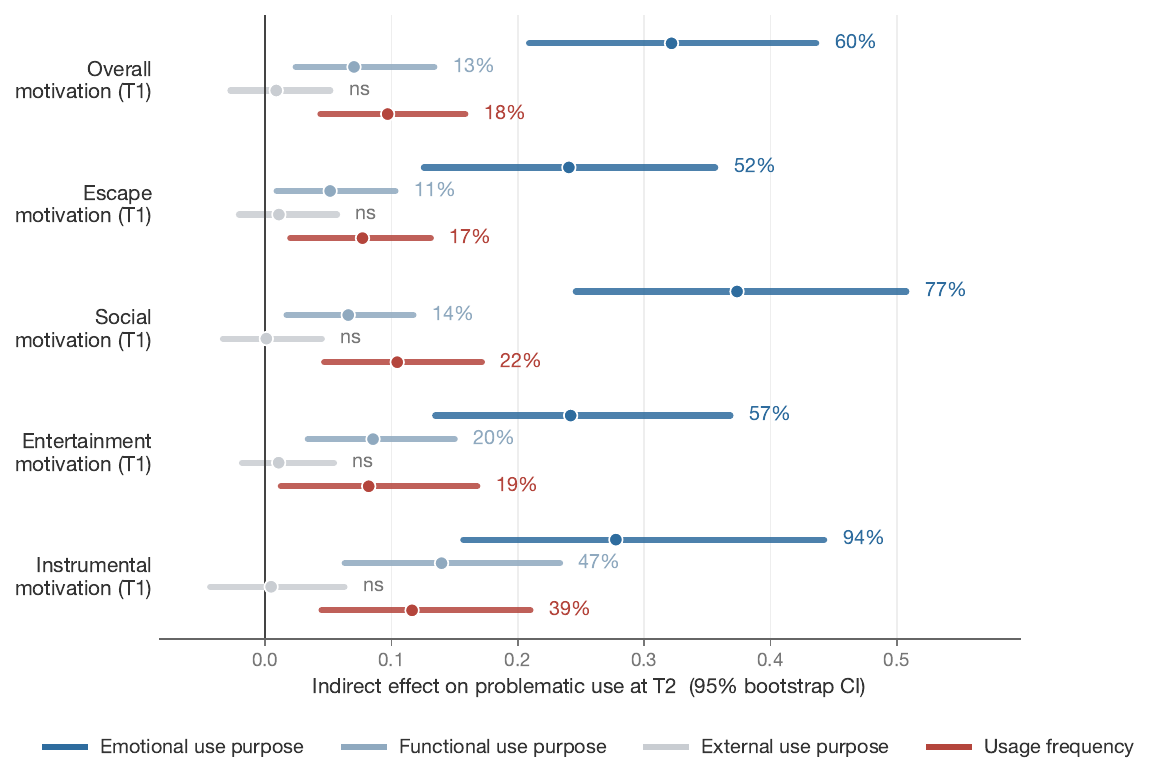}
\caption{Indirect effect of each T1 motivation measure on problematic AI use at T2 through each T2 mediator, from single-mediator bootstrap models with age, education, and baseline mental-health status as covariates. Bars indicate 95\% confidence intervals ($N$=1000). Colors indicate emotional use purpose (dark blue), functional purpose (light blue), external purpose (gray), and usage frequency (red). Channels whose interval includes zero are marked ns.}
\label{fig:mediation}
\end{figure}

On the other hand, emotional use purpose was found to account for most of the association between motivation and problematic use. 
We formally tested three use purpose scales (emotional, functional, and external) along with the usage frequency in a single-mediator model (Table~\ref{tab:mediation}, Figure~\ref{fig:mediation}).
For overall motivation, the indirect effect through emotional use purpose was .32, 95\% CI [.21, .44], against .10 [.04, .16] through usage frequency and .07 [.02, .13] through functional purpose.
The same ordering held for every motivation subscale.

With emotional purpose as the mediator in the model, the direct effect fell to non-significance for social ($c'$=.11, $p$=.30) and instrumental motivation ($c'$=.02, $p$=.77) and dropped by half or more for the rest; with usage frequency in the model, every direct effect remained significant.
In other words, emotional use purpose accounted for 52\% of the total effect for escape motivation and 77\% for social motivation, against 17\% and 22\% respectively for usage frequency. 
Emotional use purpose mediated for all four motivation types, while functional purpose carried a smaller mediation.
External use purpose mediated nothing for any motivation type. 

Finally, we ran a parallel model with four mediators simultaneously, so that each indirect effect can be estimated with the other three held constant (Appendix~\ref{appx:results:parallel}).
Emotional use purpose remained the largest channel for every motivation measure (specific indirect .21--.32), whereas functional use purpose no longer mediated at all.
Usage frequency was attenuated to roughly half its single-mediator size but remained significant (.03--.05). 
In summary, we found external purpose to mediate nothing, functional purpose to mediate consistently less than emotional purpose, and usage frequency remained a smaller but reliable channel throughout.

\begin{table}[h]
\centering
\small
\caption{Single-mediator models (95\% confidence intervals; $N$=1000): each T1 motivation measured through each T2 mediator to problematic AI use at T2, with age, education, and baseline mental-health status as covariates in both stages. $a$ = motivation$\rightarrow$mediator; $b$ = mediator$\rightarrow$outcome; $c'$ = direct effect with the mediator included. \% med.\ is the indirect effect as a share of the total, shown only where the indirect interval excludes zero; it is a derived ratio that is unstable at this sample size and is reported for ordering, not as an effect size. Mediator and outcome are measured in the same wave. $\dagger$ $p$<.10; * $p$<.05; ** $p$<.01; *** $p$<.001.}
\label{tab:mediation}
\begin{tabular}{llccclc}
\textbf{Motivation (T1)} & \textbf{Mediator (T2)} & \textbf{$a$} & \textbf{$b$} & \textbf{$c'$} & \textbf{Indirect [95\% CI]} & \textbf{\% med.} \\
\hline
Overall & Emotional purpose & .58 & .55 & .22* & .32 [.21, .44] & 60\% \\
Overall & Usage frequency & .33 & .29 & .44*** & .10 [.04, .16] & 18\% \\
Overall & Functional purpose & .37 & .19 & .47*** & .07 [.02, .13] & 13\% \\
Overall & External purpose & .05 & .18 & .53*** & .01 [$-$.03, .05] ns & --- \\
\hline
Escape & Emotional purpose & .41 & .59 & .22** & .24 [.13, .36] & 52\% \\
Escape & Usage frequency & .22 & .36 & .38*** & .08 [.02, .13] & 17\% \\
Escape & Functional purpose & .17 & .30 & .41*** & .05 [.01, .10] & 11\% \\
Escape & External purpose & .06 & .18 & .45*** & .01 [$-$.02, .06] ns & --- \\
\hline
Social & Emotional purpose & .60 & .62 & .11 ns & .37 [.25, .51] & 77\% \\
Social & Usage frequency & .33 & .32 & .38*** & .10 [.05, .17] & 22\% \\
Social & Functional purpose & .25 & .27 & .42*** & .07 [.02, .12] & 14\% \\
Social & External purpose & .00 & .20 & .48*** & .00 [$-$.03, .04] ns & --- \\
\hline
Entertainment & Emotional purpose & .40 & .61 & .18** & .24 [.13, .37] & 57\% \\
Entertainment & Usage frequency & .22 & .36 & .34*** & .08 [.01, .17] & 19\% \\
Entertainment & Functional purpose & .34 & .25 & .34*** & .09 [.03, .15] & 20\% \\
Entertainment & External purpose & .06 & .18 & .41*** & .01 [$-$.02, .05] ns & --- \\
\hline
Instrumental & Emotional purpose & .41 & .67 & .02 ns & .28 [.16, .44] & 94\% \\
Instrumental & Usage frequency & .30 & .39 & .18* & .12 [.04, .21] & 39\% \\
Instrumental & Functional purpose & .47 & .30 & .16$\dagger$ & .14 [.06, .23] & 47\% \\
Instrumental & External purpose & .02 & .20 & .29*** & .00 [$-$.04, .06] ns & --- \\
\hline
\end{tabular}
\end{table}

\section{Discussion}

The goal of our study was to understand what precedes problematic use of conversational AI systems, and whether the development of problematic use is attributable by how much people use these systems or by what they use them for.
We measured people's motivation for using AI a year before problematic use was assessed.
Our study found three key findings. 
One, individual characteristics explained little of who developed problematic AI use beyond being relatively older in age (\textbf{RQ1}).
Two, we found that AI use motivation of any kind tested, as well as overall,  predicted problematic AI use one year in advance (\textbf{RQ2}).
Three, the association between motivation and problematic use operated primarily through what people used AI for rather than how much they used it, with emotional use purpose accounting for a substantially larger share of the total effect than usage frequency for every motivation subcomponent (\textbf{RQ3}).
Following our analysis results, we discuss what they might suggest about where risk accumulates, what we gain from a year of measurement, challenges we may encounter in observing these patterns, and implications for the design of AI systems.

\subsection{Risk lies in the drift toward emotional use purpose}\label{sec:discussion:drift}

Our results imply that the stereotypical way of looking only at vulnerable and lonely individuals as at-risk for addictive and problematic behaviors is likely to be inaccurate.
In our study, we did not find individual characteristics, including having a pre-existing mental health condition, significantly predicted problematic AI use. Instead, we found that high levels of motivation of any type could be useful early warning signals. Moreover, when we put all four motivation subcomponents together in our models, none retained a distinguishable coefficient. 
This indicates that, in our data, it was not the type of motivation to use AI that mattered but instead the level of motivation. 

We also found that the four motivation types tested had something in common: their relationship with later problematic use was mediated by the emotional use of AI. This was true even for instrumental motivation--most of the predictive power of this motivation type was subsumed by the measure of emotional purpose. This suggests that task-oriented or instrumental intentions for using AI is not inherently protective; people with such utilitarian goals for AI might still use it for emotional reasons. We need to look at \textit{how} people are using AI, not just \textit{why} they want to use it. 

Our study distinguishes earlier motivation from use purpose reported at follow-up. However, because use purpose was not measured at baseline, repeated measurements are needed to determine whether individuals' use actually shifts toward emotional purposes over time.
This aligns with the process-based description of problematic technology use, where the predisposition operates through the specific behavioral pattern it produces rather than the disposition alone~\cite{brand2019interaction}, as well as earlier studies where the risk is attributable to the function that the technology serves rather than the exposure to it~\cite{kardefelt2017conceptualizing}.

We did find it surprising that there were no residual effects of the 2024 experimental study~\cite{chandra2025longitudinal} on either motivation levels or AI use purposes between the baseline usage and active usage groups. Perhaps the intervention was not strong enough to generate long-lasting differences in AI engagement, even though the active usage group was asked to use AI for social-emotional purposes for five weeks. This does imply that assigning certain types of AI uses may not lead to an increased risk for problematic usage.

\subsection{Implications for design and measurement}\label{sec:discussion:implications}

Based on our findings about where the risk of problematic AI use might be anticipated, we make two key points for implications. 

\paragraph{Implication 1: Ask what people are seeking, not only how much they use}

We found that usage frequency indeed predict problematic AI use, and it remained a reliable channel even when the other three competing mediators are held constant. 
But it made a minor contribution to the overall association between motivation and problematic use. 
We highlight this to set appropriate expectations for the often-used deployment telemetry measure.
What our study found to matter more is what a user is seeking from a system (motivation) and what they end up using it for (emotional use purpose). 
Motivation may be invisible in logs and must be asked to the users. 
On the other hand, while it may not be perfect, use purpose can be inferred from conversational logs \cite{phang2025investigating}, and measuring the shifts in use purpose can go a long way in predicting problematic use.

\paragraph{Implication 2: Watch for drift toward emotional use rather than profiling users in advance}

We found that many of the individual characteristics besides age explained very little of who developed problematic AI use. 
This finding is interpreted as inconclusive because nine predictors for the small sample size ($n$=114) cannot detect small effects. 
Together with our other finding, if risk accumulates through what use becomes rather than through who begins it, screening users at sign-up may be looking in the wrong place. 
As mentioned, attending to how their use changes may be the right place.
What our study does not address is the risk introduced by the surrounding environment, such as workplace, platforms, and social norms. 
Understanding how external factors influence people toward more AI use is critical for designing interventions for the development of problematic AI use. 
If external factors are substantial, asking people to self-manage their use is largely inadequate, as they are putting a burden of regulation on users rather than the design and environment that cultivates engagement.
Given our findings that risk is reachable from any starting point, future research should also investigate how systems are designed to influence that drift.

\subsection{Limitations and Future Directions}\label{sec:limitations}

Our study has limitations, which also suggest interesting directions for future research. 
First, we saw that problematic AI use was uncommon in our sample. 
Because this was in alignment with what was observed in the development of the source instrument~\cite{yu2024development}, we attribute this to be an artifact of sampling a general population rather than a clinical population. 
This means that our estimates describe variation within a largely non-problematic population and cannot make claims about those whose use become severe. 
Because the retained cohort has fewer number of heavier baseline users and those with mental health conditions, our retained sample is at even lower risk compared to the original cohort. 
Therefore, our finding should be interpreted as a conservative estimate. We provide additional rank-based estimation in Appendix~\ref{appx:results:rank} to further support our findings, but our estimates for a low-risk sample still cannot make claims about a high-risk sample.
Future work should deliberately recruit larger samples that include heavier AI users and people who report difficulty controlling their use or resulting interference with everyday life.
Such samples would make it possible to test whether the associations observed here extend across different levels of severity.

Second, problematic AI use has no measurement before T2 in our data, which constrained the design of our study. 
Because the candidate mediators were observed in the same wave as the outcome, the mediation models can only provide a descriptive decomposition of the association rather than establish a causal sequence~\cite{maxwell2007bias}.
The cohort's mental health worsened over the study year: 25 participants reporting a condition at follow-up who had not reported one at intake, and only two reported they no longer had a mental health condition. 
Our study cannot explain what caused these moves as these transitions afford very little statistic power. 
Neither problematic AI use nor usage was associated with the self-reported mental health condition at follow-up.
However, our mental health measure was a single self-reported item. 
Our findings can only speak to a pattern of engagement with AI rather than a clinical outcome. 
Future longitudinal studies can measure motivation, use purpose, usage frequency, and problematic use repeatedly across several waves.
Such designs could establish the temporal ordering of these constructs and examine when and for whom emotional use develops or changes.

Another limitation concerns our overall AI-use motivation composite. 
Prior work conceptualizes AI-use motivation through four distinct dimensions: \textit{escape, social, entertainment, and instrumental}, and does not specify an overall score~\cite{huang2024ai}. 
In the absence of a validated global measure, we averaged the four subscales' means to provide an equally weighted summary of ``overall.'' 
This study-defined composite should not be interpreted as showing that the dimensions are interchangeable or form a single additive construct. 
Its prospective association with problematic use suggests that broad motivational intensity carries signal, but our sample did not permit reliable distinctions among motivation types. 
Social motivation was nevertheless most strongly associated with subsequent emotional use purpose, consistent with prior evidence that social and escape motivations may be particularly relevant to AI dependence~\cite{huang2024ai}. 
Future work should validate a parsimonious general motivation measure while continuing to examine the four dimensions separately.

All associations and estimations are within the participant pool recruited in prior work~\cite{chandra2025longitudinal} among participants who consented to participate in this follow-up study, and we cannot make generalizable claims about other populations. 
Finally, the AI landscape changed substantially and quickly over the study year; we cannot separate change in the systems from the changes observed in the people using them. 

\section{Conclusion}\label{sec:Conclusion}
As conversational AI becomes embedded in everyday life, we may see more psychological impacts over time, as we have with computer-based internet, gaming, and social media use. Our findings imply that we must look beyond how much people are using these systems to why they are turning to them, and how they are using them, to accurately predict problematic AI use. We must also avoid blaming personal characteristics as the culprit for dysfunction; our study did not find strong evidence that certain types of people are vulnerable, even those who were explicitly instructed to use AI for social and emotional purposes during the prior study. Instead, high motivation and high emotional use served as stronger warning signals in our study. For HCI, this shifts attention from monitoring usage volume or profiling who might be at risk toward understanding the evolution of motivation and behavior. As AI systems become more engaging, designers should consider not only how much they encourage people to use AI, but whether they are encouraging motivations and emotional uses that may be associated with problematic use over time.



\bibliographystyle{ACM-Reference-Format}
\bibliography{_references}

\appendix
\renewcommand{\thetable}{A\arabic{table}}
\renewcommand{\thefigure}{A\arabic{figure}}
\setcounter{table}{0}
\setcounter{figure}{0}

\section{Survey Instruments and Measures}\label{appx:methods:measures}

\subsection{Instrument inventory}\label{appx:methods:unanalyzed}\label{appx:methods:items}

Table~\ref{appx:methods:tab-inventory} lists every instrument administered in the study surveys, the waves at which it was collected, its source, and whether it enters this paper's analyses.

\begin{table}[H]
\centering\small
\caption{Instruments administered across the three waves. $^{a}$Introduced or adapted in this paper.}
\label{appx:methods:tab-inventory}
\begin{tabular}{p{6.2cm}lll}
\hline
Instrument & Waves & Source & In analyses \\
\hline
Motivation for AI use & T0, T1, T2 & \cite{huang2024ai} & Yes \\
AI dependence & T0, T1, T2 & \cite{huang2024ai} & Yes \\
Problematic AI use$^{a}$ & T2 & adapted from \cite{yu2024development} & Yes \\
AI use purpose$^{a}$ & T2 & adapted from \cite{wolf2024chatgpt} & Yes \\
Usage frequency$^{a}$ & T0, T1, T2 & this study & Yes \\
General attitudes toward AI (GAAIS) & T0, T1, T2 & \cite{schepman2020gaais} & Yes \\
Social connectedness & T2 & \cite{lee1995connectedness} & Yes \\
Loneliness (three-item) & T0, T2 & \cite{hughes2004loneliness} & Yes \\
Trait impulsivity (I-8) & T0 & \cite{groskurth2022impulsive} & Yes \\
Mental-health condition & T0, T2 & self-report & Yes \\
Demographics (age, gender, race, education, household size) & T0, T2 & self-report & Yes \\
Liking People Scale & T0, T1, T2 & \cite{filsinger1981liking} & No \\
Perceived negative impacts of AI & T1, T2 & this study & No \\
Open-ended usage prompts & T2 & this study & No \\
\hline
\end{tabular}
\end{table}

\subsection{Scales introduced in this paper}\label{appx:methods:ouritems}

The following scales are introduced or adapted in this paper. Here we report the items as administered, the response options, and the scoring rule.
Attention-check items are shown where they were embedded and are excluded from all composites.

\paragraph{Problematic AI use.}
Adapted from~\citeauthor{yu2024development}.
Ten items, administered at T2 only, rated on a 5-point agreement scale (1 = Strongly Disagree, 2 = Somewhat Disagree, 3 = Neutral, 4 = Somewhat Agree, 5 = Strongly Agree); the composite is the mean of the ten items (Table~\ref{appx:methods:tab-items-pu}).
Two source-scale items were not administered (anxiety-and-irritability when unable to access; attempted to reduce usage without success), each because it overlaps with an AI dependence item administered in the same survey. Item 6 is added by our study.
Appendix~\ref{appx:methods:sumscore} reports a sensitivity analysis on the nine source-scale items alone.

\begin{table}[H]
\centering\small
\caption{Problematic AI use: items as administered (T2 only).}
\label{appx:methods:tab-items-pu}
\begin{tabular}{p{12.5cm}}
\hline
1. I constantly have thoughts related to AI conversational agents lingering in my mind. \\
2. I frequently find myself opening AI conversational agents even when I had no initial intention to use one. \\
3. I progressively spend more time with AI conversational agents. \\
4. I have lost interest in previously enjoyable activities due to using AI conversational agents. \\
5. My use of AI conversational agents causes me to procrastinate and delay completing necessary tasks. \\
6. I become fully absorbed when using AI conversational agents. \\
7. I devote excessive time to AI conversational agents despite problems. \\
8. I suffer from sleep deprivation due to excessive use of AI conversational agents. \\
9. I hide the extent of my AI conversational agent usage from family, friends, or therapists. \\
10. I turn to AI conversational agents to alleviate feelings of helplessness or anxiety. \\
\hline
\end{tabular}
\end{table}

\paragraph{AI use purpose.}
Adapted from Wolf and Maier~\cite{wolf2024chatgpt}.
Fourteen scored items plus one embedded attention check, administered at T2 only, with the stem ``I use AI conversational agents\ldots'' rated on a 5-point agreement scale (1 = Strongly Disagree, 2 = Somewhat Disagree, 3 = Neutral, 4 = Somewhat Agree, 5 = Strongly Agree).
Each scale is the mean of its items. 
One item is reverse-scored, and one is not assigned to a subscale due to ambiguity (Table~\ref{appx:methods:tab-items-purpose}).

\begin{table}[H]
\centering\small
\caption{AI use purpose items and scale assignment (T2 only). (R) = reverse-scored.}
\label{appx:methods:tab-items-purpose}
\begin{tabular}{p{10.5cm}l}
\hline
Item & Scale \\
\hline
To overcome certain difficulties. & Emotional \\
To enhance my learning and knowledge. & Functional \\
For the pleasure it gives me to know more about it. & (unassigned) \\
To manage my feelings. & Emotional \\
Please select neutral for this statement. & (attention check) \\
Because I am encouraged or required to. & External \\
Because I am told not to use it. (R) & External \\
Because many people I know use it. & External \\
Because people I respect use it. & External \\
To look cool. & External \\
Because it helps me to accomplish tasks more quickly. & Functional \\
To improve my performance. & Functional \\
Increase my productivity. & Functional \\
To make my life easier. & Functional \\
Because I find it helpful. & Functional \\
\hline
\end{tabular}
\end{table}

\paragraph{Usage frequency.}\label{appx:methods:platforms}
Study-constructed matrix item.
At T2, participants rated ``In general, how frequently do you use the listed artificial intelligence (AI) conversational agent platforms?'' per platform on an 8-point frequency scale (1 = Never, 2 = Less than once per month, 3 = Once per month, 4 = Multiple times per month, 5 = Once per week, 6 = Multiple times per week, 7 = Once per day, 8 = Multiple times per day) across 13 platforms: Microsoft Copilot, Google Gemini, ChatGPT, Pi AI, Perplexity, Claude, Deepseek, Snapchat AI, Grok, Replika, Character.AI, Meta AI, and other AI conversational agents.
The analysis variable is the \emph{maximum} across the 13 platforms; a sum across platforms is not used because it conflates intensity of use with breadth of adoption.
At T0 and T1, usage was measured with a 5-platform matrix on a daily frequency scale (1 = Once a day or never through 5 = 10+ times in a day), and the analysis variables likewise take the maximum across platforms.

\subsection{Measurement structure and reliability}\label{appx:methods:reliability}

Table~\ref{appx:methods:tab-cfa-fit} reports the confirmatory factor analyses for the motivation and use-purpose scales. 

\begin{table}[H]
\centering
\small
\caption{Confirmatory factor analyses of motivation.}
\label{appx:methods:tab-cfa-fit}
\begin{tabular}{lcccccc}
\hline
Model & $\chi^2$($df$) & $\chi^2/df$ & CFI & TLI & RMSEA & SRMR \\
\hline
Motivation 4-factor (T2, 12 items) & 78.6 (48) & 1.64 & .973 & .963 & .075 & .059 \\
Use purpose 3-factor (T2, 13 items) & 133.4 (62) & 2.15 & .899 & .873 & .101 & .073 \\
Use purpose 2-factor (Emotional+Functional, 8 items) & 65.9 (19) & 3.47 & .915 & .874 & .148 & .060 \\
\hline
\end{tabular}
\end{table}

\section{Scoring Sensitivity for the Primary Outcome}\label{appx:methods:sumscore}

Our study scores problematic AI use as the mean of the ten administered items; ~\citeauthor{yu2024development} report the source scale as a sum~\cite{yu2024development}.
Every participant answered all ten items, so the sum is an exact linear transform of the mean (sum = $10\times$mean).
Table~\ref{appx:methods:tab-sumscore} compares our distribution on the summed metric with the source instrument's published distribution.
A second variant restricts the composite to the nine source-scale items, dropping the investigator-added item~6 (Table~\ref{appx:methods:tab-items-pu}).
Results are unchanged: every T1 motivation measure remains significant (escape $\beta$=.30, social $\beta$=.32, instrumental $\beta$=.17, entertainment $\beta$=.25, overall $\beta$=.34; all $p$<.001).

\begin{table}[H]
\centering
\small
\caption{Problematic AI use on the summed metric compared with the source scale's published distribution~\cite{yu2024development}. Our sums span the 10 administered items and the 9 source-scale items (dropping the investigator-added item 6) on a 5-point scale (possible ranges 10--50 and 9--45); the source spans 11 items on a 4-point scale (range 11--44). Per-item $M$ = sum divided by the number of items.}
\label{appx:methods:tab-sumscore}
\begin{tabular}{lccccccc}
\hline
 & $M$ & Per-item $M$ & $SD$ & Median & 80th pct. & 90th pct. & Skewness \\
\hline
This study (10-item sum) & 16.2 & 1.62 & 6.8 & 14 & 21 & 26 & 1.39 \\
This study (9-item sum, source items only) & 13.8 & 1.53 & 5.9 & 12 & 17 & 23 & 1.61 \\
~\citeauthor{yu2024development} (11-item sum) & 15.9 & 1.44 & 6.9 & 13 & 21 & 26 & 1.71 \\
\hline
\end{tabular}
\end{table}

\section{Covariate-Set Sensitivity}\label{appx:methods:controls}

Table~\ref{appx:methods:tab-controls} re-estimates the primary motivation models under four different covariate specifications: no controls, the primary set (age, education, baseline mental-health status), the primary set plus baseline usage, and a set screened on correlation with the outcome (age, T2 loneliness, baseline usage).
All twenty motivation coefficients remain significant after correction under every specification.

\begin{table}[H]
\centering
\small
\caption{Control-specification sensitivity for the T1 motivation models ($n$=114). *** $p$<.001.}
\label{appx:methods:tab-controls}
\begin{tabular}{lcccc}
\hline
Motivation (T1) & No controls & Primary & + baseline usage & Correlation-screened \\
\hline
Overall & 0.373*** & 0.367*** & 0.359*** & 0.335*** \\
Escape & 0.290*** & 0.312*** & 0.298*** & 0.281*** \\
Social & 0.343*** & 0.328*** & 0.315*** & 0.282*** \\
Entertainment & 0.288*** & 0.287*** & 0.271*** & 0.242*** \\
Instrumental & 0.207*** & 0.201*** & 0.186*** & 0.200*** \\
\hline
\end{tabular}
\end{table}

\section{Experimental Arm Differences}\label{appx:methods:arms}

Table~\ref{appx:methods:tab-arms} summarizes the differences observed between two experimental arms.
The only nominal difference is loneliness ($p$=.024 unadjusted), with prior Active Use arm participants reporting higher loneliness.
As a further check on pooling the arms, we re-estimated the five primary motivation models with arm assignment added as a covariate, and the motivation coefficients are essentially unchanged ($\beta$=.20--.36, all $p$<.001, versus $\beta$=.20--.37 without), and the assignment coefficient is not significant in any model (all $p\geq.17$).

\begin{table}[H]
\centering
\small
\caption{Experimental arm differences at T2 (Welch two-sample $t$-tests; AU arm $n$=69 vs.\ BU arm $n$=45). Positive $d$ indicates that AU arm is higher. $q$ from Benjamini--Hochberg correction within this 8-test family; 0/8 survive.}
\label{appx:methods:tab-arms}
\begin{tabular}{lcccccc}
\hline
Outcome (T2) & $M$ treat (AU) & $M$ control (BU) & $t$ & $p$ & $d$ & $q$ \\
\hline
Positive AI attitudes (T2) & 3.71 & 3.55 & 1.03 & .308 & .19 & .410 \\
Negative AI attitudes (T2) & 3.21 & 3.11 & 0.59 & .556 & .12 & .635 \\
AI dependence (T2) & 7.06 & 6.58 & 1.05 & .295 & .20 & .410 \\
Problematic use (T2) & 1.68 & 1.52 & 1.22 & .225 & .23 & .410 \\
Social motivation (T2) & 1.70 & 1.50 & 1.20 & .232 & .22 & .410 \\
Escape motivation (T2) & 1.48 & 1.44 & 0.32 & .748 & .06 & .748 \\
Social connectedness (T2) & 3.46 & 3.69 & -1.03 & .304 & -.20 & .410 \\
Loneliness (T2) & 6.38 & 5.60 & 2.28 & .024 & .43 & .196 \\
\hline
\end{tabular}
\end{table}

\section{Motivation Models at T0}\label{appx:methods:t0}

Table~\ref{appx:methods:tab-t0} reports the same model specification as the present study, but with motivation measured at T0. The pattern matches the T1 models.

\begin{table}[H]
\centering
\small
\caption{Motivation measured at T0 as a predictor of problematic use at T2 ($n$=114). ** $p$<.01; *** $p$<.001.}
\label{appx:methods:tab-t0}
\begin{tabular}{lcc}
\hline
Motivation (T0) & $\beta$ & Adjusted $R^2$ \\
\hline
Overall & 0.346*** & .341 \\
Escape & 0.323*** & .316 \\
Social & 0.289*** & .268 \\
Entertainment & 0.233*** & .207 \\
Instrumental & 0.163** & .146 \\
\hline
\end{tabular}
\end{table}

\section{Distributional Robustness of the Primary Models}\label{appx:results:rank}

Table~\ref{appx:results:tab-rank} re-estimates the five primary T1 motivation models under three alternative treatments of the outcome's distribution: (1) a rank-transformed specification (outcome and motivation predictor replaced by ranks), (2) a log-transformed outcome, and (3) a logistic model of scoring above 2.0 on the 1--5 scale.
Every motivation measure remains significant under every specification, and the rank-based coefficients are larger than the OLS estimates, so the reported estimates are conservative with respect to the floor.

\begin{table}[H]
\centering
\small
\caption{Distributional robustness of the primary T1 motivation models ($n$=114), with age, education, and baseline mental-health status as covariates. Rank = OLS on rank-transformed outcome and predictor; log = log-transformed outcome; logistic = $P$(problematic use > 2.0), $n$=24 above the cut. All estimates significant (largest $p$=.018).}
\label{appx:results:tab-rank}
\begin{tabular}{lcccc}
\hline
Motivation (T1) & $\beta$ OLS (reported) & $\beta$ rank & $\beta$ log & OR logistic \\
\hline
Overall & .367 & .564 & .206 & 2.89 \\
Escape & .312 & .463 & .172 & 1.87 \\
Social & .328 & .479 & .179 & 2.56 \\
Instrumental & .201 & .353 & .121 & 2.40 \\
Entertainment & .287 & .468 & .164 & 2.52 \\
\hline
\end{tabular}
\end{table}

\section{Motivation Subscale Distributions and Coefficient Equality}\label{appx:results:jointmodel}

Table~\ref{appx:results:tab-distributions} reports the floor and ceiling shares and standard deviations of the T1 motivation subscales, and Table~\ref{appx:results:tab-equality} reports the coefficient-equality tests in the joint model entering all four subscales.
Floor means that participants answered Strongly Disagree to all three of its items (subscale mean = 1.0, the scale minimum), and ceiling means that they answered Strongly Agree to all three (mean = 4.0).
Motivation subscales are not statistically distinguishable.

\begin{table}[H]
\centering
\small
\caption{Distributions of the T1 motivation subscales ($n$=114).}
\label{appx:results:tab-distributions}
\begin{tabular}{lccc}
\hline
Subscale & \% at floor & \% at ceiling & SD \\
\hline
Escape & 52 & 4 & 0.93 \\
Social & 45 & 4 & 0.91 \\
Instrumental & 2 & 38 & 0.65 \\
Entertainment & 12 & 5 & 0.89 \\
\hline
\end{tabular}
\end{table}

\begin{table}[H]
\centering
\small
\caption{Coefficient-equality tests among the four T1 motivation subscales in a joint model of problematic use at T2 ($n$=114; Wald tests on the difference of standardized coefficients, HC3).}
\label{appx:results:tab-equality}
\begin{tabular}{llcc}
\hline
Test & Comparison & Statistic & $p$ \\
\hline
pairwise & Escape vs Social & -0.11 & .914 \\
pairwise & Escape vs Instrumental & 0.89 & .372 \\
pairwise & Escape vs Entertainment & 0.45 & .650 \\
pairwise & Social vs Instrumental & 0.91 & .362 \\
pairwise & Social vs Entertainment & 0.64 & .521 \\
pairwise & Instrumental vs Entertainment & -0.31 & .753 \\
omnibus (all four equal) & all & 0.63 & .597 \\
\hline
\end{tabular}
\end{table}

\section{Individual Characteristics as Predictors of Problematic AI Use}\label{appx:results:characteristics}

Table~\ref{appx:results:tab-characteristics} reports the joint model behind \textbf{RQ1}: problematic AI use at T2 regressed on the full set of individual characteristics in one OLS model.
Age is the only predictor that survives correction; race is nominally significant but does not.

\begin{table}[H]
\centering
\small
\caption{Individual characteristics as joint predictors of problematic AI use at T2 ($n$=114; one OLS model, standardized continuous predictors, HC3 standard errors; gender, mental-health condition, and race entered as 0/1). Adjusted $R^2$ = .137. $q$ from Benjamini--Hochberg correction within this 9-test family.}
\label{appx:results:tab-characteristics}
\begin{tabular}{lccccc}
\hline
Characteristic & $\beta$ & $SE$ & 95\% CI & $p$ & $q$ \\
\hline
Age & .236 & .062 & [.115, .357] & $<$.001 & .001 \\
Education & .054 & .068 & [$-$.080, .188] & .427 & .702 \\
Gender (female) & .014 & .137 & [$-$.254, .282] & .920 & .920 \\
Mental-health condition (T0) & .210 & .133 & [$-$.052, .471] & .116 & .347 \\
Loneliness (T0) & .010 & .071 & [$-$.129, .148] & .891 & .920 \\
Impulsivity (T0) & .050 & .069 & [$-$.085, .185] & .468 & .702 \\
Race (White, T0) & $-$.329 & .160 & [$-$.644, $-$.015] & .040 & .180 \\
Household size (T0) & .021 & .084 & [$-$.143, .185] & .804 & .920 \\
Baseline usage (T0) & .096 & .073 & [$-$.046, .238] & .185 & .417 \\
\hline
\end{tabular}
\end{table}

\section{Individual Characteristics as Predictors of Motivation}\label{appx:results:who}

Table~\ref{appx:results:tab-who} reports the individual characteristics set as joint predictors of overall motivation at T1, at T2, and of its T1 to T2 change.
No coefficient survives correction in any model.

\begin{table}[H]
\centering
\small
\caption{Individual characteristics as joint predictors of overall motivation ($n$=114; one OLS model per column). Adjusted $R^2$ = .059 (T1), .089 (T2), .025 (T1 to T2 change).}
\label{appx:results:tab-who}
\begin{tabular}{lccc}
\hline
Characteristic & $\beta$ (T1) & $\beta$ (T2) & $\beta$ (change) \\
\hline
Age & .066 & .132 & .066 \\
Education & -.082 & -.027 & .055 \\
Gender (female) & -.120 & -.173 & -.052 \\
Mental-health condition (T0) & -.007 & .045 & .052 \\
Loneliness (T0) & -.021 & .039 & .060 \\
Impulsivity (T0) & .097 & .054 & -.043 \\
Race (White, T0) & -.067 & -.297 & -.230 \\
Household size (T0) & -.006 & .011 & .017 \\
Baseline usage (T0) & .176 & .104 & -.073 \\
\hline
\end{tabular}
\end{table}

\section{Attenuation by Usage Frequency}\label{appx:results:attenuation}

Table~\ref{appx:results:tab-attenuation} reports each T1 motivation coefficient before and after T2 usage frequency enters the model.

\begin{table}[H]
\centering
\small
\caption{Attenuation of the T1 motivation coefficients when T2 usage frequency enters the model ($n$=114). All motivation coefficients remain significant after correction.}
\label{appx:results:tab-attenuation}
\begin{tabular}{lccc}
\hline
Motivation (T1) & $\beta$ solo & $\beta$ with frequency & \% removed \\
\hline
Overall & .367 & .301 & 18\% \\
Escape & .312 & .260 & 17\% \\
Social & .328 & .257 & 22\% \\
Instrumental & .201 & .122 & 39\% \\
Entertainment & .287 & .231 & 19\% \\
\hline
\end{tabular}
\end{table}

\section{Parallel Mediation and Mediator Specificity}\label{appx:results:parallel}

Table~\ref{appx:results:tab-parallel} reports the parallel model entering all four candidate mediators simultaneously, so each specific indirect effect controls the other three channels, together with bootstrap contrasts between the emotional-purpose channel and each rival.
All fifteen emotional-minus-rival contrasts exclude zero.

Because the emotional-purpose items could partially overlap with the outcome's mood-regulation item (i.e., item 10), we also re-estimated the emotional-purpose mediation against a version of the problematic AI use scale without its mood item and with each emotional-purpose item alone as the mediator.
All indirect effects remained significant (outcome without mood item: indirect effects .22--.35; single-item mediators: .14--.40).

\begin{table}[H]
\centering
\small
\caption{Parallel four-mediator model ($n$=114): all four T2 mediators entered simultaneously; specific indirect effects (each channel controlling the other three) and pairwise emotional-minus-rival contrasts, 95\% bootstrap CIs ($N$=1000). EMO = emotional use purpose; FREQ = usage frequency; FUNC = functional use purpose; EXT = external use purpose. * = CI excludes zero.}
\label{appx:results:tab-parallel}
\begin{tabular}{llll}
\hline
Motivation (T1) & Quantity & Estimate [95\% CI] & \\
\hline
Overall & specific indirect: EMO & .28 [.17, .41] & * \\
Overall & specific indirect: FREQ & .05 [.01, .10] & * \\
Overall & specific indirect: FUNC & -.03 [-.08, .03] & ns \\
Overall & specific indirect: EXT & .01 [-.01, .04] & ns \\
Overall & contrast EMO - FREQ & .23 [.11, .37] & * \\
Overall & contrast EMO - FUNC & .31 [.18, .46] & * \\
Overall & contrast EMO - EXT & .28 [.17, .40] & * \\
\hline
Escape & specific indirect: EMO & .21 [.11, .34] & * \\
Escape & specific indirect: FREQ & .03 [.00, .07] & * \\
Escape & specific indirect: FUNC & -.01 [-.04, .02] & ns \\
Escape & specific indirect: EXT & .01 [-.01, .03] & ns \\
Escape & contrast EMO - FREQ & .18 [.08, .30] & * \\
Escape & contrast EMO - FUNC & .22 [.11, .36] & * \\
Escape & contrast EMO - EXT & .20 [.10, .33] & * \\
\hline
Social & specific indirect: EMO & .32 [.20, .48] & * \\
Social & specific indirect: FREQ & .05 [.00, .11] & * \\
Social & specific indirect: FUNC & -.01 [-.05, .03] & ns \\
Social & specific indirect: EXT & .00 [-.02, .03] & ns \\
Social & contrast EMO - FREQ & .28 [.14, .44] & * \\
Social & contrast EMO - FUNC & .33 [.19, .51] & * \\
Social & contrast EMO - EXT & .32 [.21, .47] & * \\
\hline
Entertainment & specific indirect: EMO & .22 [.11, .34] & * \\
Entertainment & specific indirect: FREQ & .04 [.00, .09] & * \\
Entertainment & specific indirect: FUNC & -.03 [-.10, .02] & ns \\
Entertainment & specific indirect: EXT & .01 [-.01, .04] & ns \\
Entertainment & contrast EMO - FREQ & .18 [.09, .29] & * \\
Entertainment & contrast EMO - FUNC & .25 [.12, .41] & * \\
Entertainment & contrast EMO - EXT & .21 [.10, .33] & * \\
\hline
Instrumental & specific indirect: EMO & .25 [.13, .40] & * \\
Instrumental & specific indirect: FREQ & .05 [.00, .11] & * \\
Instrumental & specific indirect: FUNC & -.03 [-.12, .06] & ns \\
Instrumental & specific indirect: EXT & .00 [-.02, .04] & ns \\
Instrumental & contrast EMO - FREQ & .20 [.09, .34] & * \\
Instrumental & contrast EMO - FUNC & .28 [.12, .49] & * \\
Instrumental & contrast EMO - EXT & .25 [.13, .39] & * \\
\hline
\end{tabular}
\end{table}

\section{Temporal Shifts in Dependence, Motivation, and Attitudes}\label{appx:discussion:shifts}

\begin{figure}[H]
\centering
\includegraphics[width=0.9\linewidth]{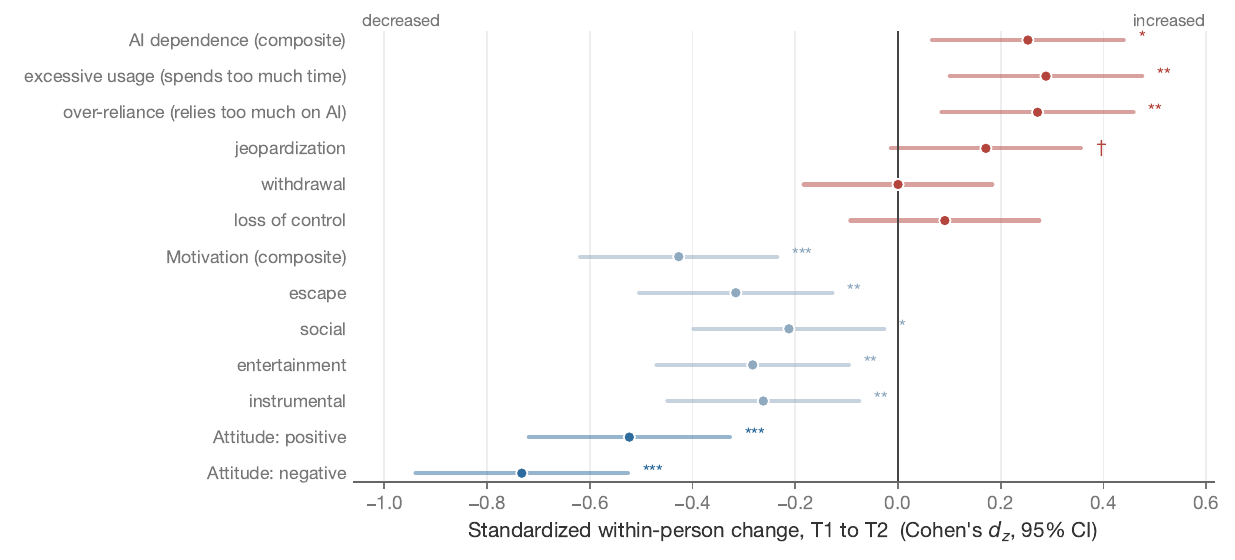}
\caption{Standardized within-person change from T1 to T2 (Cohen's $d_z$ with 95\% confidence intervals, $n$=114). Self-assessed AI dependence and its excessive-usage and over-reliance facets rose, while stated motivation (all four subscales and the composite) and both positive and negative attitudes toward AI declined. Stars reflect $q$ (Benjamini--Hochberg within the shift family): $\dagger$ $q$<.10; * $q$<.05; ** $q$<.01; *** $q$<.001.}
\label{appx:discussion:fig-shifts}
\end{figure}

These individual differences and changes we observed played out against a moving population backdrop.
During the period of over one year, self-assessed AI dependence rose while attitudes toward AI---both positive and negative---declined. 
In other words, as a whole, this sample population was reporting more dependence while their feelings toward AI dissipated over time.
Because between-person differences in motivation remained stable through this period even as sample means shifted, this divergence does not undermine our prospective findings.
However, it raises a question about calibrating risk detection if the general attitude toward AI continues to become subdued. 
Table~\ref{appx:discussion:tab-shifts} reports the paired within-person shifts behind the divergence we observed in the data.
Figure~\ref{appx:discussion:fig-shifts} displays the follow-up-year (T1 to T2) shifts, including the five dependence facets.

\begin{table}[H]
\centering
\small
\caption{Paired within-person shifts in AI dependence, overall motivation, and AI attitudes ($n$=114; paired $t$-tests, $d_z$ = mean change / SD of change, $q$ = Benjamini--Hochberg within the shift family). Dependence is the 5--20 sum composite; motivation is on its 1--4 scale; attitudes on their 1--5 scale. $\%\uparrow$ = share of participants whose score increased.}
\label{appx:discussion:tab-shifts}
\begin{tabular}{llccrrcc}
\hline
Measure & Interval & $M$ before & $M$ after & $\Delta$ & $d_z$ & $q$ & $\%\uparrow$ \\
\hline
AI dependence & T0 to T1 & 6.01 & 6.33 & +0.32 & .18 & .084 & 32 \\
AI dependence & T1 to T2 & 6.33 & 6.87 & +0.54 & .25 & .011 & 39 \\
AI dependence & T0 to T2 & 6.01 & 6.87 & +0.86 & .39 & $<$.001 & 41 \\
\hline
Overall motivation & T0 to T1 & 2.13 & 2.36 & +0.23 & .44 & $<$.001 & 61 \\
Overall motivation & T1 to T2 & 2.36 & 2.17 & -0.19 & -.43 & $<$.001 & 29 \\
Overall motivation & T0 to T2 & 2.13 & 2.17 & +0.04 & .09 & .445 & 52 \\
\hline
Positive attitudes & T0 to T1 & 3.84 & 3.90 & +0.06 & .15 & .125 & 45 \\
Positive attitudes & T1 to T2 & 3.90 & 3.65 & -0.25 & -.52 & $<$.001 & 30 \\
Positive attitudes & T0 to T2 & 3.84 & 3.65 & -0.20 & -.38 & $<$.001 & 34 \\
\hline
Negative attitudes & T0 to T1 & 3.45 & 3.54 & +0.10 & .21 & .040 & 48 \\
Negative attitudes & T1 to T2 & 3.54 & 3.17 & -0.38 & -.73 & $<$.001 & 18 \\
Negative attitudes & T0 to T2 & 3.45 & 3.17 & -0.28 & -.50 & $<$.001 & 23 \\
\hline
\end{tabular}
\end{table}

\end{document}